\documentclass[fleqn,twoside]{article}
\usepackage{espcrc2}

\usepackage{graphicx}

\usepackage{amssymb}

\usepackage[figuresright]{rotating}

\def\hc#1{\leavevmode\hbox to\columnwidth{\hss #1\hss}\leavevmode}

\newcommand{\AmS}{{\protect\the\textfont2
  A\kern-.1667em\lower.5ex\hbox{M}\kern-.125emS}}

\def\degr{\hbox{$^\circ$}}
\def\arcmin{\hbox{$^\prime$}}
\def\arcsec{\hbox{$^{\prime\prime}$}}
\def\utw{\smash{\rlap{\lower5pt\hbox{$\sim$}}}}
\def\udtw{\smash{\rlap{\lower6pt\hbox{$\approx$}}}}

\def\fdg{\hbox{$.\!\!^\circ$}}
\def\farcm{\hbox{$.\mkern-4mu^\prime$}}
\def\farcs{\hbox{$.\!\!^{\prime\prime}$}}

\def\diameter{{\ifmmode\mathchoice
{\ooalign{\hfil\hbox{$\displaystyle/$}\hfil\crcr
{\hbox{$\displaystyle\mathchar"20D$}}}}
{\ooalign{\hfil\hbox{$\textstyle/$}\hfil\crcr
{\hbox{$\textstyle\mathchar"20D$}}}}
{\ooalign{\hfil\hbox{$\scriptstyle/$}\hfil\crcr
{\hbox{$\scriptstyle\mathchar"20D$}}}}
{\ooalign{\hfil\hbox{$\scriptscriptstyle/$}\hfil\crcr
{\hbox{$\scriptscriptstyle\mathchar"20D$}}}}
\else{\ooalign{\hfil/\hfil\crcr\mathhexbox20D}}%
\fi}}

\newif\ifForReferee \ForRefereefalse
\ifForReferee
\makeatletter
\def\small{\@setfontsize\small\@xipt{16}%{13.6}%
\abovedisplayskip 11\p@ \@plus3\p@ minus6\p@
\belowdisplayskip \abovedisplayskip
\abovedisplayshortskip  \z@ \@plus3\p@
\belowdisplayshortskip  6.5\p@ \@plus3.5\p@ minus3\p@
\def\@listi{\leftmargin\leftmargini
 \parsep 4.5\p@ \@plus2\p@ minus\p@ \itemsep \parsep
            \topsep 9\p@ \@plus3\p@ minus5\p@}}
\makeatother
\fi
\title{The Technical Performance of the HEGRA System of Imaging Air Cherenkov Telescopes}

\author{
G.~P{\"u}hlhofer\address[MPIK]{Max-Planck-Institut f{\"u}r Kernphysik, Postfach 103980, D-69029 Heidelberg, Germany}\address[CORRAUTH]{Corresponding author: G. P{\"u}hlhofer, e-mail Gerd.Puehlhofer@mpi-hd.mpg.de},
O.~Bolz\addressmark[MPIK],
N.~G\"otting\address[UNIHH]{Universit\"at Hamburg, Institut f\"ur Experimentalphysik, Luruper Chaussee 149, D-22761 Hamburg, Germany},
A.~Heusler\addressmark[MPIK],
D.~Horns\addressmark[MPIK],
A.~Kohnle\addressmark[MPIK],
H.~Lampeitl\addressmark[MPIK],
M.~Panter\addressmark[MPIK],
M.~Tluczykont\addressmark[UNIHH],
F.~Aharonian\addressmark[MPIK],
A.~Akhperjanian\address[YER]{Yerevan Physics Institute, Alikhanian Br. 2, 375036 Yerevan, Armenia},
M.~Beilicke\addressmark[UNIHH],
K.~Bernl\"ohr\addressmark[MPIK],
H.~B\"orst\address[UNIKIEL]{Universit\"at Kiel, Institut f\"ur Experimentelle und Angewandte Physik, Leibnizstra{\ss}e 15-19, D-24118 Kiel, Germany},
H.~Bojahr\address[UNIWUP]{Universit\"at Wuppertal, Fachbereich Physik, Gau{\ss}str.20, D-42097 Wuppertal, Germany},
T.~Coarasa\address[MPIMU]{Max-Planck-Institut f\"ur Physik, F\"ohringer Ring 6, D-80805 M\"unchen, Germany},
J.L.~Contreras\addressmark[MPIMU],
J.~Cortina\address[BARC]{Now at Institut de F\'{\i}sica d'Altes Energies, UAB, Edifici Cn, E-08193, Bellaterra (Barcelona), Spain},
S.~Denninghoff\addressmark[MPIMU],
M.V.~Fonseca\address[MAD]{Universidad Complutense, Facultad de Ciencias F\'{\i}sicas, Ciudad Universitaria, E-28040 Madrid, Spain},
M.~Girma\addressmark[MPIK],
G.~Heinzelmann\addressmark[UNIHH],
G.~Hermann\addressmark[MPIK],
W.~Hofmann\addressmark[MPIK],
I.~Jung\addressmark[MPIK],
R.~Kankanyan\addressmark[MPIK],
M.~Kestel\addressmark[MPIMU],
A.~Konopelko\addressmark[MPIK],
H.~Kornmeyer\addressmark[MPIMU],
D.~Kranich\addressmark[MPIMU],
M.~Lopez\addressmark[MAD],
E.~Lorenz\addressmark[MPIMU],
F.~Lucarelli\addressmark[MAD],
O.~Mang\addressmark[UNIKIEL],
H.~Meyer\addressmark[UNIWUP],
R.~Mirzoyan\addressmark[MPIMU],
A.~Moralejo\addressmark[MAD],
E.~Ona-Wilhelmi\addressmark[MAD],
A.~Plyasheshnikov\addressmark[MPIK]\address[ALTAI]{On leave from Altai State University, Dimitrov Street 66, 656099 Barnaul, Russia},
R.~de\,los\,Reyes\addressmark[MAD],
W.~Rhode\addressmark[UNIWUP],
J.~Ripken\addressmark[UNIHH],
G.~Rowell\addressmark[MPIK],
V.~Sahakian\addressmark[YER],
M.~Samorski\addressmark[UNIKIEL],
M.~Schilling\addressmark[UNIKIEL],
M.~Siems\addressmark[UNIKIEL],
D.~Sobzynska\addressmark[MPIMU]\address[LODZ]{Home institute: University Lodz, Poland},
W.~Stamm\addressmark[UNIKIEL],
V.~Vitale\addressmark[MPIMU],
H.J.~V\"olk\addressmark[MPIK],
C.~A.~Wiedner\addressmark[MPIK],
W.~Wittek\addressmark[MPIMU]\\
\vspace{1ex}HEGRA Collaboration
}

\begin{document}

\thispagestyle{empty}

\begin{abstract}
\hrule
\vskip 0.5pc
{\noindent \bf Abstract}

\noindent Between early 1997 and late 2002, the HEGRA collaboration operated a
stereoscopic system of 4 (later 5) imaging atmospheric Cherenkov telescopes.
%on La Palma in the Canary Islands. 
In this paper we present the calibration
schemes which were developed for the system, and report on the performance
of the detector over the years.
In general, the telescope system was very well understood, regarding both the absolute calibration
and the slight changes in performance over the years.
The system had an energy threshold of 500\,GeV for observations at zenith
and under optimum detector conditions.
With the corresponding calibration schemes, a systematic accuracy of 15 percent on the absolute
energy scale has been achieved.
The continuous sensitivity monitoring provided a relative
accuracy of a few percent, and showed that the threshold
did not exceed 600\,GeV throughout the entire operation time.
The readout electronics and the imaging quality of the dishes were well monitored and stable. 
The absolute pointing had an accuracy of at least 25\arcsec; this number was guaranteed throughout the
whole lifetime of the experiment.
% for espcrc2

\vskip 0.5pc
\hrule
\vskip 0.5pc
{\noindent \sl Preprint submitted to Elsevier Science\hfill\today}
\end{abstract}

% for elsart
%\begin{keyword}
%Imaging air Cherenkov technique \sep Very high energy gamma ray astronomy
%\PACS 95.55.Ka \sep 95.55.Vj \sep 96.40.Pq
%\end{keyword}

% for elsart
%\end{frontmatter}

\maketitle

\section{Introduction}

The HEGRA\footnote{HEGRA stands for "High Energy Gamma Ray Astronomy"}
stereoscopic Imaging Atmospheric Cherenkov Telescope (IACT) system was located 
in the Canary Islands, 
at 2200\,m above sea level on the Roque de los Muchachos on La Palma (17\degr52\arcmin34\arcsec\ West,
28\degr45\arcmin34\arcsec\ North).
It consisted of 5 identical telescopes (CT\,2 - CT\,6), which operated in coincidence for the
stereoscopic detection of air showers in the atmosphere.
The HEGRA collaboration also ran a single telescope (CT\,1 \cite{MirzoyanCT1,Mrk501PaperII})
in stand-alone mode, which is not covered in this paper.

The atmospheric Cherenkov technique uses the
Cherenkov light emitted by the charged shower particles, in order
%The goal of such an experiment is
to reconstruct arrival direction and energy of the primary TeV photons
that initiated the showers.
At the same time, a large background consisting
of showers induced by the
isotropically incoming charged cosmic ray particles is detected.
Since the TeV photons point back to their sources, information
on the sites which emit TeV radiation is directly obtained.

HEGRA has published a large number of detailed analyses of characteristics of Galactic and extragalactic 
sources, some of them containing data sets which were collected over several years. 
Analysis techniques and the performance of the IACT system 
as well as Monte Carlo simulations have been described in detail
in \cite{DaumPerformance,Mrk501PaperI,Mrk501TimeAveraged,HEGRAPerformance99}. 
%The scope of this paper is to describe the long-term calibration and stability of the
%telescope system throughout several years of operation.
Now, after the disassembly of the system, 
we felt compelled to provide a detailed account of the long-term performance
and stability of the system throughout several years of operation.
Data for this topic have been collected mostly via the standard calibration procedure, some
information was gathered using special calibration setups.
Earlier results, based on data which were recorded until the end of 2000,
%accounting for the status of the experiment until the year 2000, 
were in parts already presented in \cite{HEGRAPerformanceICRC2001}.

The calibration of the telescopes comprises the following major issues:

{\bf Continuous camera electronics calibration:} The relative gain of the camera pixels 
          and the pixel timing was calibrated by so called {\it laser runs}, 
	  in which the whole camera was illuminated
	  uniformly by laser flashes. These runs were performed every night.

{\bf Pointing calibration:} The telescopes' pointing was calibrated and monitored by 
	  dedicated calibration runs, so called {\it point runs}, which used
	  stars as reference objects. Those calibration runs were typically performed every few months.

{\bf Imaging properties:} The quality of the optical point spread function ({\it psf}) of the dishes 
          was monitored using the same {\it point runs}. 
%%%          To a certain extent, continuous monitoring 
%%%          was also provided by the measurement of the cosmic ray background events.
          The influence of the {\it psf} on $\gamma$-ray induced shower images could best be  
          obtained from regular observations
%	  of the strong TeV $\gamma$-ray sources Crab, Mkn\,501 and Mkn\,421. 
	  of strong TeV $\gamma$-ray sources.
%	  namely the standard candle Crab with constant flux, 
%	  and the flaring sources Mkn\,501 and Mkn\,421 with occasional high flux states. 

{\bf Relative sensitivity monitoring:} Continuous
	  monitoring of changes in the light sensitivity of the complete detectors,
	  including mirror reflectivity and quantum efficiency of
	  the PMs, could be obtained by the measurement of the event trigger rate which originates
	  from the steady flux of cosmic rays. 
	  Additionally, the absolute gain of the camera electronics alone was
          monitored by the same {\it laser runs} as mentioned above. 

{\bf Absolute energy threshold:} The absolute light sensitivity of the telescopes,
          and hence the energy threshold of the system, was addressed by so called
	  {\it muon runs} which were performed roughly once or twice per year,
          and also by special laser setups using a calibrated reference photodiode.

These topics are covered in more detail in this paper, after an overview on the
telescopes' hardware and an introduction to the data processing and general analysis techniques
in the next two sections.

\section{The telescope system hardware}

\begin{figure}[t]
%\vspace*{2.0mm}
%\begin{center}
%\epsfig{file=telescope,width=\hsize}
\hc{\includegraphics*[width=1.0\columnwidth]{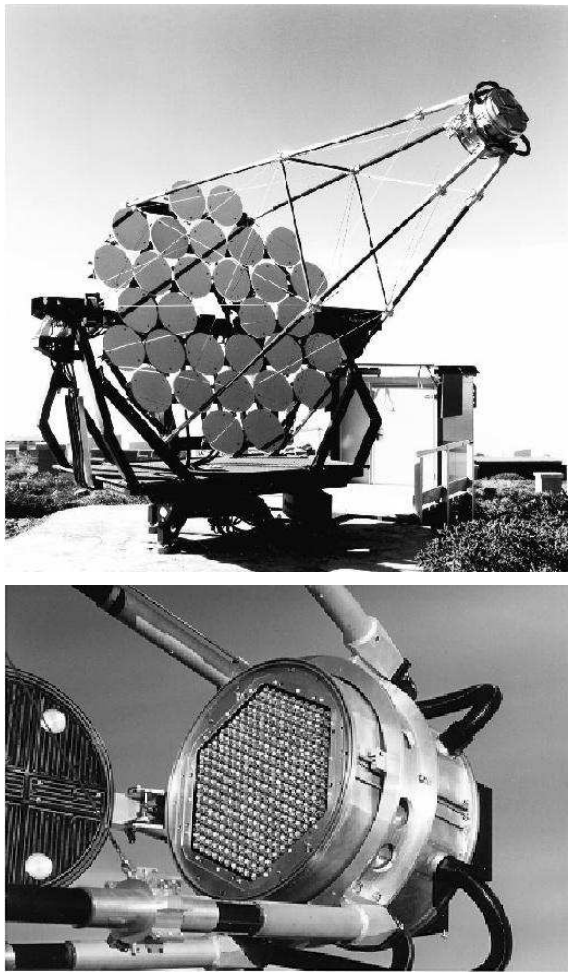}}
\caption{Up: One of the system telescopes. The readout electronics is located in a container
         which can be seen in the background of the telescope.
	 Down: Front view of the camera; 
         a funnel plate in front of the hexagonally arranged pixel matrix collects the
	 incoming light onto the photocathodes.}
\label{F:telescope}
%\end{center}
\end{figure}

Figure~\ref{F:telescope} shows one system telescope and a front view of its camera.
Each telescope has an altitude-azimuth mount. The reflector has a diameter of $3.40\,\mathrm{m}$,
with a focal length of $4.92\,\mathrm{m}$.
The mirror is segmented, each dish holds 30 spherical glass mirrors with diameters of 60\,cm;
the total mirror area per telescope is $8.5\,\mathrm{m}^{2}$.
The mirrors are aluminized and quartz-coated. The mirror segments are arranged 
in a so called Davies-Cotton design \cite{DaviesCotton}, resulting in a design point spread function
having a spot size of better than $3\arcmin$ 
(width of a two-dimensional Gaussian distribution) on the
optical axis and not worse than $4\arcmin$ at the edge of the field of view.  

The telescopes are driven by stepper motors with a step size of $1\farcs3$. 
%%% 14-bit shaft encoders ($1\farcm3$ per bit) are used to measure the absolute position of the telescope axes. 
The absolute position of the telescope axes are measured using 14-bit shaft encoders ($1\farcm3$ per bit).
The time used for tracking is provided by the local telescope CPUs
which are synchronized to a GPS clock. The accuracy with which the CPU clocks are synchronized 
is one second. 
The online tracking accuracy is better than $0\fdg1$ in most cases, being limited by
the mechanical distortions of the telescope's structures. 
After application of offline corrections which account for these effects, 
the pointing accuracy is better than $25\arcsec$ \cite{HEGRAPointing}.

\begin{figure}[t]
%\begin{center}
\hc{\includegraphics[width=1.0\columnwidth]{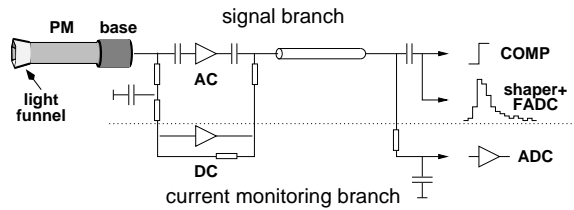}}
\caption{Schematic overview of the processing of the electric signals from a camera pixel.}
\label{F:signals}
%\end{center}
\end{figure}

The camera in the focal plane consists of 271 pixels, which are arranged in a hexagonal close
packed layout. Each pixel contains a photomultiplier (PM) (EMI 9083 KFLA) and a preamplifier
located in the base.
A metallized funnel plate in the focal plane
concentrates the incoming photons onto the photocathodes; the funnels are shaped 
similar to Winston cones.
The distance between the centers of two adjacent pixels is $\approx 21\,\mathrm{mm}$,
corresponding to $0\fdg2445$ at the sky.
The total field of view covers a hexagon with side length of $20\,\mathrm{cm}$, 
equivalent in area to a circle of $4\fdg3$ diameter.
%Each pixel views $0\fdg2445$ at the sky, the total field of view of a camera is $4\fdg3$;  
Cables provide high voltage (HV) supply from the electronics to the
camera and signal transfer back. For each single pixel, the HV can be adjusted separately
within a range of 285\,V below a camera-wide maximum HV value; 
the pixel values across all cameras range from 600 to 1000\,V.

Both AC and DC content of the PM signal are amplified separately in the camera, and then
transferred to the readout electronics (see
Fig.\,\ref{F:signals}), which is housed by a container a few meters away from each telescope.   
Except for the HV supply and the next neighbour trigger module,
all electronics are built in VME standard; the VME bus is
controlled by a 68040 Motorola CPU.
The fast pulses induced by air shower events are digitized with 120\,MHz, 8-bit flash ADCs (FADC);
the pulses are shaped with a time constant of $\tau=12\,\mathrm{ns}$ in order to match the
time resolution of the FADCs. The DC currents of the pixels are measured for monitoring purposes:
During data taking, pixels with a DC current above $3\,\mathrm{\mu A}$ (basically because of stars
traversing the 
field of view) are excluded from the trigger decision to avoid an excessive level of random triggers.
These pixels are also excluded from offline image analysis.
Besides that, the current measurement is useful for the pointing calibration (see \S\,\ref{S:pointing}).
   
The trigger decision for an event is made in two stages \cite{HEGRATriggerPaper}:
A local trigger is released when two 
adjacent pixels exceed a certain threshold. The comparator thresholds were initially
set to 10\,mV, later on to 8\,mV, corresponding to 8 and 6 photoelectrons (ph.e.), respectively.
The ``next-neighbour''-decision is done hard-wired 
with a coincidence window of 12\,ns; the hardware solution provides fast triggers
and allows a low pixel threshold.
The cameras thus trigger with a rate of the order of $10^{3} - 10^{4}\,\mathrm{Hz}$.
All of these local triggers are sent to a central station.
If at least two telescopes have fired within 70\,ns
(corrected for the zenith angle dependent propagation delay),
an event trigger is sent back to all telescopes, with a frequency of typically 15\,Hz.
Here, the continuous signal digitization by the FADCs is stopped at all telescopes,
including those without a local trigger.
%For each event, 16 FADC bins in the buffers, corresponding to 128\,ns,
%which contain the event pulses, and the information which pixels have triggered are read out. 
For each event, the FADCs and the information on which pixels have triggered are read out.
From the FADC buffers, a window of 16 bins per channel (corresponding to 128\,ns)
containing the event pulse is extracted.
After a zero-suppression, 
the event data are then sent to the central processing computer which performs online event
building and stores the data to disk. 
The event time is determined from a GPS time receiver, with an accuracy of $200\,\mathrm{ns}$;
for check reasons and possibly as fall-back,
also the data from a local rubidium clock is read out and stored \cite{AndreasCrabundGeminga}.
The dead time after an event is $200\,\mu\mathrm{s}$ (non-extendable, meaning that
further events triggering during that readout time do not prolong the dead time interval). 

The first four system telescopes (CT\,3 - CT\,6) were installed during 1995 and 1996 \cite{DaumPerformance}. 
The system was running in the final hardware layout
%a stable configuration 
between March 1997 and September 2002. Major changes during that time
were the reduction of the pixel
threshold from 8 photoelectrons (ph.e.) to the final value of 6\,ph.e.\ in May 1997,
and the upgrade of the CT\,2
electronics and its inclusion into the system as fifth telescope in August 1998; however,
technical problems with CT\,2 allowed the operation of a complete stable 5-telescope system 
only since June 1999.

\section{Data processing, analysis techniques and performance}
\label{S:dataprocessing}

During data taking, various preliminary analysis steps were performed, including
an online event display and a fast search for $\gamma$-ray signals 
after each data run of typically 20 or 30 minutes.
For final processing, the data were brought
via tape from La Palma to a home institute at the end of an observation period (3 weeks).
%At the end of an observation period (3 weeks), the
%data were brought via tape from La Palma to a home institute for further data processing. 
In the following, the processing of a normal data run is described:

\subsection{Pulse shape analysis (pixel level)} 

        For each event, the FADC pulses have to be transformed into pixel amplitudes.
        First, the zero offsets of the FADCs (pedestals) of each pixel are subtracted;
	the pedestals are determined dynamically with a 
	frequency of a few
        seconds, from the first two FADC bins which, by design,
	do not contain any shower information.
        A deconvolution function is applied to restore the original pulse. 
        Then the time jitter between individual pixels (few ns) is corrected,
        based on calibration results which are discussed in \S\,\ref{S:camera}. 
        An event time is determined from all active pixels above a certain threshold. 
%	The FADC window of each pixel is reduced to 5 bins around that event time. 
	Finally, pixel amplitudes $c_{pix}$ are computed from two 
	FADC bins around that event time.
        Pixels with overflow values in the FADCs are treated with a different procedure.
	Details 
%%%     on the procedure 
        are described in \cite{MarkusThesis} and \cite{HEGRAPaperOnShowerProfiles}.

\subsection{Image processing (single telescope level)}
\label{SS:imageprocessing}

\begin{description}
\item[Pixel processing:]
            The pixel amplitudes are transformed into calibrated values, 
	    in units of the number of photoelectrons (ph.e.) which the PM's photocathode has emitted.
            For that purpose, all pixel amplitudes are multiplied with their respective
            relative gain factors, which are determined from the preceding {\it laser run}
            (camera flatfielding),
            and a camera-wide conversion factor $\kappa_{\mathrm{el}}$; the
	    latter is determined on a monthly basis.
            Also, a non-linear response function for high amplitudes above $\sim 150\,\mathrm{ph.e.}$ is applied. 
            Pixels which are defective or have a DC current above a certain threshold are set to zero.
            Details of this procedure are discussed in \S\,\ref{S:camera} (camera electronics) and in
	    \S\,\ref{SS:conversion} (conversion factor).
\item[Image parameters:]
            To distinguish the shower image in the camera from noise pixels, a so called two-level tail
	    cut 
            is applied: all pixels above 6\,ph.e., and all pixels above 3\,ph.e.\ which have an adjacent
            pixel above 6\,ph.e.\ are accepted.
            From the pixel amplitudes, 
            the following Hillas parameters \cite{HillasParameter} are computed for each telescope:
	    center of gravity of the image $\vec{x}$,
            orientation of the major image axis $\vartheta$,
            width $w$, length $l$ and the image amplitude $amp$ (sometimes called image size).
            Also, the amplitudes of the two ``hottest'' pixels are stored, 
            i.e.\ the pixels with the highest amplitudes which have (most probably) caused the local
            telescope trigger. These values can be used e.g.\ to track the nonlinear behaviour of the
	    readout chain at high amplitudes, or to apply a software threshold similar to
	    the hardware trigger condition (cf.\ \S\,\ref{SS:relativecheck}).
\end{description}

        An event data record finally consists of the
        image parameters of all telescopes, the event time, and the current actual orientation of each
	telescope; these values are written to a DST (''Data Summary Table'') file.
        For most analysis purposes, only DST data are required, which provide a fast and easy
	access to the data.

\subsection{Shower reconstruction (combining data from all telescopes)}

  From the image data, the shower geometry, the shape of the shower, and the energy of the
  primary particle are reconstructed.
  Here, we concentrate on the following standard analysis chain
  in conjunction with its relevant calibration issues:

\subsubsection{Direction and core location}
\label{SSS:directionandcore}
        The shower axis in space is obtained from a pure stereoscopic reconstruction; different 
        algorithms are described in \cite{HofmannStereoTechniques}.
        The direction of the incoming primary particle,
        for $\gamma$-rays equivalent to the $\gamma$-ray source location,
        is directly obtained, since at TeV energies flight direction of the primary and the shower
        axis are practically identical.

        Before the shower axis is reconstructed by any stereoscopic algorithm,
        the positions of the image centers $\vec{x}_{tel}$ are corrected for the
	mispointings of the individual telescopes. Details of this procedure
	are discussed in \S\,\ref{S:pointing}.
        After these corrections, all cameras view exactly the same sky sector. The shower direction
	is computed directly in this sky coordinate frame, onto which the shower images are 
	projected by the telescope imaging, by determining the point in the sky where all projected
	shower axes are crossing. 

        The reconstructed shower directions are then used to produce a sky map.
        For point sources, usually
	the squared angular distance $\theta^{2}$ of the shower direction to the source location
	in the field of view (FoV) is plotted. In this representation, a pure
	background distribution is expected to be flat, while $\gamma$-ray sources
	produce an excess at zero.
	All events below an angular cut $\theta_{\mathrm{max}}$ are counted as ON
	source events. The background estimate is usually derived from 
        other areas $A$ in the FoV, and the normalisation factor 
        $\alpha_{\mathrm{ON/OFF}} = A_{\mathrm{ON}} / A_{\mathrm{OFF}}$ 
	reflects the different opening angles for ON and OFF source
	counting in the FoV.

        The intersection point of the shower axis with the ground 
        (i.e.\ the projected impact point of the shower on ground) is called the shower core.
        The shower core is computed similarly to the shower direction,
        in a projection of the images
 	onto a plane at ground level which is parallel to the system pointing.
        For advanced energy reconstruction algorithms, the height of the shower maximum
        is determined in addition \cite{HofmannImprovedEnergy}.

        The reconstructed core locations are used in the subsequent shape and energy calculations.
        Note that because of the stereoscopic view of a shower with a telescope system,
        the geometric reconstruction does not depend on the results 
        of those following analysis steps.

\begin{figure}[t]
%\begin{center}
\hc{\includegraphics[width=1.0\columnwidth]{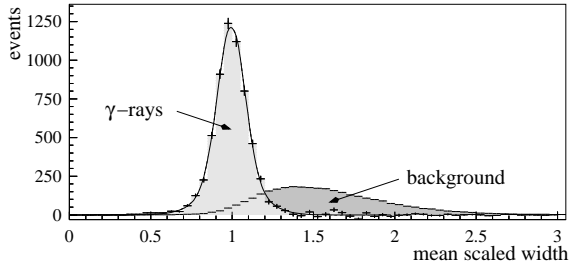}}
\caption{Distribution of the image parameter {\it mean scaled width}:
         The left, light-shaded histogram shows $\gamma$-rays, derived from 
         background-subtracted on-source observations on Mkn\,501.
         The distribution can be described by a Gaussian with a mean of 1.0 and a width of 0.1;
	 a better fit can be obtained by a double-Gaussian (solid line).
         The right, dark-shaded histogram shows background events, i.e.\ mostly charged
	 cosmic rays, derived from control positions in the FoV; 
	 the distribution was normalized to the on-source area which was
	 used for the $\gamma$-ray events.}
\label{F:meanscaledwidth}
%\end{center}
\end{figure}

%  \item[Primary particle identification:]
\subsubsection{Primary particle identification by the shape parameter {\em mean scaled width}}
        Particle identification -- in most cases on statistical grounds -- is provided by the
        difference of the shower development between $\gamma$-rays and charged cosmic rays (CR),
        which induce a permanent isotropic background. Usually, a simple quantifier
        like the so called {\it mean scaled width} ({\it msw}) \cite{Mrk501PaperI} is defined
        to provide information how $\gamma$-ray-like an event was.

%        its value is
%        calculated from measured event parameters and 
%        expectation values derived from Monte Carlo simulations,
%        making use of the reconstructed distance of the shower core to the
%        telescopes.

        The {\it mean scaled width} is defined as the average of
	all {\it scaled width} ($sw_{tel}$) values \cite{Mrk501PaperI}
        \begin{equation}
           msw      = \frac{1}{N_{tel}} \sum_{tel} sw_{tel}
        \end{equation}
        with
        \begin{equation}
           sw_{tel} = \frac{w_{tel}}{w_{\mathrm{exp}}(amp_{tel},dtc_{tel},zen)},
        \end{equation}
%        \begin{equation}
%           msw = \left(\sum_{tel}\right)^{-1}
%                 \sum_{tel} w_{tel} w^{-1}_{\mathrm{exp}}(amp_{tel},dtc_{tel},zen)
%        \end{equation}
        where $w_{\mathrm{exp}}$ is the expected value for $w$ in a $\gamma$-ray event,
        given the image amplitude $amp_{tel}$, the core distance from the telescope $dtc_{tel}$,
        and the zenith angle $zen$. The sum runs over all $N_{tel}$ telescopes 
        which have an amplitude $amp_{tel}$ greater than 40\,ph.e. 
        To enhance a $\gamma$-ray signal, events with a $msw$ value above a certain threshold
%        $msw_{\mathrm{max}}$ 
        (typically 1.1 or 1.2) are rejected (see Fig.\,\ref{F:meanscaledwidth}), 
	the corresponding cut efficiencies are given in Table\,\ref{T:hegra_systemparameter}.

        The expectation values $w_{\mathrm{exp}}$ 
	were initially derived from Monte Carlo simulations 
        \cite{HEGRAPerformance99}, which perform full shower and detector simulations.
%	and were also calibrated using real CR data.
        Monitoring and fine-tuning of the values for
        $w_{\mathrm{exp}}$ are discussed in \S\,\ref{S:pointspread}.

%        The {\it mean scaled width}
%        has proven to be very robust under varying detector conditions and provides 
%        good signal to noise ratios. 

%  \item[Energy reconstruction:]
\subsubsection{Energy reconstruction}
\label{SSS:energyreco}
        An energy estimate, presuming that the event was induced by a
        $\gamma$-ray, is similarly computed from measured event parameters and 
        expectation values derived from Monte Carlos
        \cite{Mrk501PaperI,Mrk501TimeAveraged,HEGRAPerformance99,HofmannImprovedEnergy,HofmannMrk501Reanalyse}.   
%	, again using the core position

        In the standard approach, at first an energy estimate $E_{tel}$ is derived for each individual telescope.
        The values are derived from a lookup table
        \begin{equation}
%           f:amp^{*}_{tel},dtc_{tel},zen \rightarrow E_{tel}
           E_{tel} = f(amp^{*}_{tel},dtc_{tel},zen),
        \end{equation}
        where $dtc_{tel}$ is the core distance, $zen$ the zenith angle, and $amp^{*}_{tel}$
	is a rescaled value of the measured image amplitude $amp_{tel}$.
        The values for the lookup table were derived from Monte Carlo simulations.
        The simulations were adjusted according to the status of the experiment in summer 1997. 
        Subsequent changes in the light sensitivity of the detectors are accounted for by 
        rescaling the measured amplitude values $amp_{tel}$ by a factor $\kappa'_{\mathrm{opt}}$;
        the rescaled values are labeled
	$amp^{*}_{tel}$. More details on the procedure are given in \S\,\ref{S:relative}.

        The energy estimate $E$ of an event is then calculated as
%       \cite{Mrk501PaperI}
        \begin{equation}
           E = \exp \left( \frac{1}{N_{tel}} \sum_{tel}\ln \left(E_{tel}\right) \right).
        \end{equation}
        The sum runs over all $N_{tel}$ telescopes which have an amplitude
	$amp^{*}_{tel}$ greater than $40\,\mathrm{ph.e.^{*}}$ and are not too far away
	from the core location ($dtc_{tel} < 200\,\mathrm{m}$ for zenith angles below 50\degr).

%        For a given core distance $dtc_{tel}$ and zenith angle $zen$, 
%        the relation between energy $E_{tel}$ and
%        image amplitude is monotonous. Therefore, one can directly derive a differential energy
%        spectrum. 
	For the spectral evaluation, only events are allowed 
	for which at least two telescopes are included in the energy estimate, 	 
	and where at least two telescopes have seen the core under angles which
	differ by more than 20\degr.
	With the given parameters, 
	a 20\% single event energy resolution is achieved, almost
	independent of the primary $\gamma$-ray energy
        \cite{Mrk501PaperI,Mrk501TimeAveraged,HofmannMrk501Reanalyse}.

% \item[Spectral evaluation:]
\subsubsection{Spectral evaluation}
\label{SSS:spectraleval}
        To derive the source energy spectrum, the events are appropriately weighted 
	and filled into a histogram. The weight function contains two pieces of information:
        (a) the so-called effective area $A_{\mathrm{eff}}(E)$, which reflects the
	energy dependent acceptance function of the instrument, and
	(b) the high level analysis cut	efficiency $\kappa_{\gamma}(E)$, which is imposed 
        by the mean scaled width cut and the angular cut.
        The differential photon flux per energy bin $E_{i}$ is then derived according to \cite{Mrk501PaperI}
	\begin{eqnarray}
           \label{G:photonflux}
           \frac{\mathrm{d}\Phi}{\mathrm{d}E}(E_{i}) = \frac{1}{\Delta t \Delta E_{i}}
           \left\{
           \sum_{j=1}^{N_{\mathrm{ON},i}}
           {\left[\kappa_{\gamma}(E_{j})A_{\mathrm{eff}}(E_{j})\right]^{-1}} \right. \nonumber \\
	   \left. \mbox{} - \alpha_{\mathrm{ON/OFF}} \sum_{j=1}^{N_{\mathrm{OFF},i}}
           {\left[\kappa_{\gamma}(E_{j})A_{\mathrm{eff}}(E_{j})\right]^{-1}}
           \right\}.
        \end{eqnarray}

        For each energy bin, the sums run over the events in the on-source and off-source areas 
        in the FoV, respectively.

        The effective area $A_{\mathrm{eff}}$ takes into account the pure trigger
	acceptance of the instrument $A'_{\mathrm{eff}}$ \cite{HEGRAPerformance99},
	calculated as
        \begin{equation}
           A'_{\mathrm{eff}}(E) = 2 \pi \int_{0}^{\infty}{P_{\gamma}(E,r)\,r\,\mbox{d}r}
        \end{equation}
        where $P_{\gamma}(E,r)$ is the trigger probability for $\gamma$-rays with a given energy
	$E$ and a core distance $r$ from the system center. 
        Moreover,
        also the finite energy resolution and low level analysis cuts need to be considered.
        This is done by integrating over the energy resolution function $p(E,\tilde{E})$, which
	denotes the probability that a true shower energy $E$ is reconstructed as $\tilde{E}$
        and includes the low level cuts
        (basically the cuts on $amp^{*}_{tel}$ and on $dtc_{tel}$, see \S\,\ref{SSS:energyreco}).
%        basically an
%        image amplitude cut of $amp_{tel} > 40\,\mathrm{ph.e.}$
%	and a core cut of $dtc_{tel}<200\,\mathrm{m}$ (for zenith angles below 50\degr).
        $A_{\mathrm{eff}}$ is then derived according to \cite{Mrk501PaperI}
        \begin{equation}
           A_{\mathrm{eff}}(\alpha;\tilde{E}) = \frac{\int{\mbox{d}E\,
           p(E,\tilde{E})\,A'_{\mathrm{eff}}(E)\,\Phi_{\alpha}(E)}}{\Phi_{\alpha}(\tilde{E})}.
        \end{equation}
%        This calculation needs an input model spectrum $\Phi_{\alpha}$
        $\Phi_{\alpha}$ is an assumed model spectrum; for many purposes the exact shape of this
	spectrum is not important.
	$\Phi_{\alpha}$ is characterized here by a spectral index $\alpha$ of an assumed power
	law; however, $\Phi_{\alpha}$ can be any model function.
	Usually one starts with
	a power law with a spectral index $\alpha$ (e.g.\ 2.5), and the final spectrum is derived
        iteratively according to the measurement \cite{Mrk501TimeAveraged}.
	
        The effective area $A_{\mathrm{eff}}$ needs to be calculated separately for
        different system setups (number of telescopes actually running in the system),
        for different trigger settings, and also
        for different trigger sensitivities due to gain changes of the system.
        The latter is done by a simple rescaling procedure;
	% rescaling $P_{\gamma}(E,r)$; 
	details are discussed in \S\,\ref{S:relative}. 

%%%        For simplicity, $A_{\mathrm{eff}}$ is used further on as a synonym for
%%%        $A'_{\mathrm{eff}}$, 
%%%        since $A_{\mathrm{eff}}$ itself is not needed any more in the following.

%        The energy spectrum is derived in the following way: First, an energy spectrum is filled,
%	where the events are weighted according to the energy dependent acceptance function of the
%	instrument, the so called effective area. The effective area accounts for the trigger
%	acceptance and low level analysis cuts, like an image amplitude cut of 40\,ph.e. and a 
%	core cut ($dtc_{tel}<200\,\mathrm{m}$ for zenith angles below 50\degr). Afterwards, each bin of the spectrum 
%        is corrected for the respective high level analysis cut efficiency, 
%	imposed by the mean scaled width cut and the angular cut.

%\end{description}

\subsection{Performance summary}

\begin{table}[t]
\begin{center}

\begin{tabular}{lll} \\ \hline \hline
  \multicolumn{2}{l}{energy threshold for $\gamma$'s}    & $500\,\mathrm{GeV}$  \\ \hline
  field of view     & \multicolumn{2}{l}{homogeneous $\gamma$-acceptance} \\
		     & \multicolumn{2}{l}{\hspace{1ex}on $\diameter \ge 2\degr$} \\ \cline{2-3}
		     & \multicolumn{2}{l}{$> 50\%$ of peak $\gamma$-acceptance} \\
		     & \multicolumn{2}{l}{\hspace{1ex}on $\diameter \simeq 4\degr$} \\ \hline
  event rate  & all events				  & 15\,Hz \\ \cline{2-3}
	       & in FoV $\diameter = 2\degr$		  & 2.3\,Hz \\ \hline
  \multicolumn{2}{l}{Crab $\gamma$-ray detection rate}   & 120 $\gamma\,\mbox{h}^{-1}$ \\ \hline
  angular	      & all events			  & $\sigma = 0\fdg09$  	 \\ \cline{2-3}
  resolution\dag     & selected events       & $0\fdg03..0\fdg12$	     \\ \hline
  shower	      & unknown source        & $\sigma \simeq 10\,\mbox{m}$ \\ 
  core\dag\dag       & position	      & 			     \\ \cline{2-3}
		      & known position        & $\sigma \simeq 3\,\mbox{m}$  \\ \hline
  energy	      & unknown source        & $\mathrm{RMS} \simeq 20\%$   \\ 
  resolution	      & position	      & 			     \\ \cline{2-3}
  \dag\dag\dag       & known position        & $\mathrm{RMS} \simeq 10\%$   \\
		      & + shower height       \\  \hline
  flux 	      & quasi BG-free,  		  & $0.3\,\mathrm{Crab}\,\times$ \\
  sensitivity        & \hspace{1ex}$t<1\,\mbox{h}$	  & \hspace{1ex} $(t/1\,\mathrm{h})^{-1}$ \\ \cline{2-3}
		      & BG dominated,			  & $0.03\,\mathrm{Crab}\,\times$  \\
		      & \hspace{1ex}$t>1\,\mbox{h}$	  & \hspace{1ex} $(t/100\,\mathrm{h})^{-\frac{1}{2}}$ \\ \hline
  background	      & loose cut		& $\kappa_{\mathrm{CR}}=14\%$,\\
   reduction I:      & \hspace{1ex}($msw<1.2$) & $\kappa_{\gamma}=96\%$\\ \cline{2-3}
   shape cut	      & tight cut		& $\kappa_{\mathrm{CR}}=7\%$,\\ 
		      & \hspace{1ex}($msw<1.1$) & $\kappa_{\gamma}=80\%$  \\ \hline
  background	      & loose cut		       & $\kappa_{\mathrm{CR}}=4.8\%$,\\
  reduction II:      & \hspace{1ex}($\theta<0\fdg22$) & $\kappa_{\gamma} > 90\%$ \\ \cline{2-3}
  direction cut      & tight cut		       & $\kappa_{\mathrm{CR}}=1.2\%$,\\ 
  point source       & \hspace{1ex}($\theta<0\fdg12$) & $\kappa_{\gamma}=60\%$ \\
  vs. $\diameter = 2\degr$ \\ \hline \hline
\end{tabular}

\end{center}
\caption{HEGRA IACT system performance for a 4 telescope setup,
         observations at zenith, and under optimum
         detector conditions. 
	 FoV: field of view. BG: background.
	 \dag: single event resolution, sigma of a 2-dimensional Gaussian
         $exp\left(-(\theta^{2}_{\mathrm{x}}+\theta^{2}_{\mathrm{y}}) / (2 \sigma^{2})\right)$. 
	 \dag\dag: single event resolution in one ground coordinate.
	 \dag\dag\dag: single event resolution, $\Delta E/E$ nearly independent of $E$.
	 The cut efficiency for $\gamma$-rays is called $\kappa_{\gamma}$, $\kappa_{\mathrm{CR}}$ is the 
	 respective fraction of background events after the cut.}
\label{T:hegra_systemparameter}
\end{table}

In Table\ \ref{T:hegra_systemparameter} we summarize the properties of the HEGRA telescope system; 
the values are given for a 4-telescope system setup and 
observations at zenith.
%%% low zenith angle observations. 
To first order, the main difference between the final 5- and the 4-telescope system
comes from an increase in the detection rate of about 15\%; since the HEGRA system has run for a
considerable time as a 4-telescope-system, most calculations and values refer to this setup.

With the HEGRA system, observations between 0\degr\ and 60\degr\ zenith angle were performed.
A detailed description of the performance dependence on zenith angle is beyond the scope of this paper.
For performance monitoring, experimental data up to different maximum angles
are used in this paper,
the limits were chosen according to the degree of the respective zenith angle dependence.
Generally, the CR background event rate changes only by a few percent 
between 0\degr\ and 30\degr. The $\gamma$-ray detection rate drops however by $\sim20\%$ over this range,
$\gamma$-ray rate comparisons therefore need a finer zenith angle binning (0\degr-20\degr, 20\degr-30\degr).

\section{Continuous camera electronics calibration}
\label{S:camera}

\begin{table}[t]
\begin{center}
\begin{tabular}{cccc} \\ \hline \hline
  year    &  mjd     &  periods  & runs           \\ \hline
  1997    &  50449-  &  56-68    & 5195-8931      \\
  1998    &  50814-  &  69-80    & 8932-13417     \\
  1999    &  51179-  &  81-92    & 13418-17661    \\
  2000    &  51544-  &  93-105   & 17662-21600    \\
  2001    &  51910-  &  106-117  & 22000-27005    \\
  2002    &  52334-  &  118-126  & 27006-29928    \\ \hline \hline \\
\end{tabular}
\end{center}
\caption{Throughout the Figures of this paper, the time development of detector parameters is given
         in one of the above listed units. Each observing period comprises three weeks around 
	 the new moon date.}
\label{T:yearperiodrunmjd}
\end{table}

%%% \begin{figure}[t]
%%% \hc{\includegraphics[width=0.65\columnwidth]{faulties.eps}}
%%% \caption{Number of defective channels per telescope as a function of the run number. The plot starts
%%% in Jan. 1997 and ends in Sept. 2002. The gap from run number 11224 to 12000 is artificial, these runs
%%% do not exist; CT\,2 was only included in the system since run 12000.
%%% Normally, only 1-2\% of all 271 channels per telescope are dead.
%%% The plots are truncated above 32 pixels; a camera with more than 15 defective 
%%% pixels is usually completely excluded from the analysis.}
%%% \label{F:faulties}
%%% \end{figure}

\begin{figure}[t]
\vspace*{2.0mm}
\hc{\includegraphics[width=1.0\columnwidth]{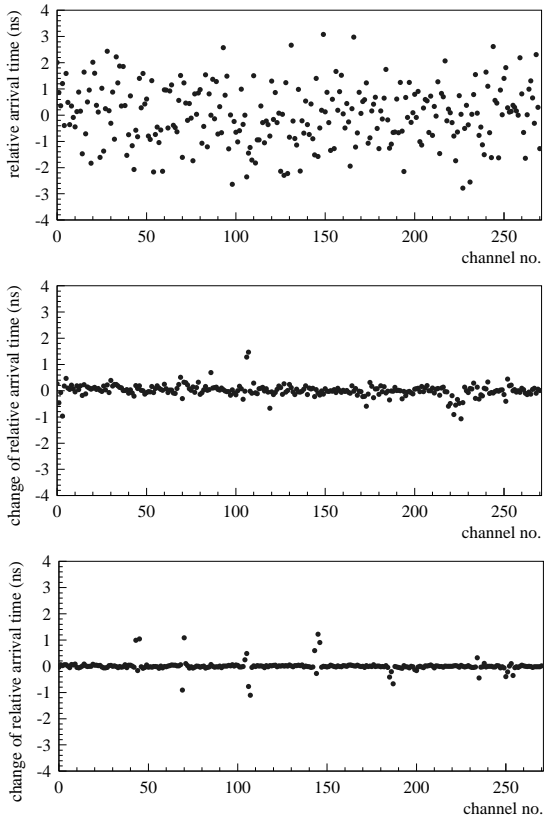}}
\caption{Results of a pixel timing calibration for one camera. 
         Upper panel: relative arrival times
         of the signals in the FADC, derived from one {\it laser run}. 
         Middle panel: change of the arrival times due to a typical HV flatfielding adjustment
	 for all camera pixels.
         Lower panel: change of the arrival times due to an exchange of two FADC modules;
         since the pixels in the camera are numbered in concentric circles,
         not all 16 channels per module are adjacent in the corresponding channel plot.}
\label{F:lasercalib}
\end{figure}

For data calibration and processing purposes, 
monitor and status information about the hardware of the cameras
%, i.e.\ the PMs and the readout electronics, 
is obtained mainly from laser calibration runs and from normal data runs.

First of all, pixel channels which are defective or found impossible to calibrate are
identified and are on a run by run basis artificially set to zero. 
Not more than 2\% of all 271 channels per telescope were excluded in most (95\%) of the runs.
%%% Figure \ref{F:faulties} shows the number of 
%%% dead channels for all system telescopes as a function of the run number.
%%% As can be seen, 
%%% only 1-2\% of all 271 channels per telescope are excluded during most of the time.
In addition, pixels which had in the last monitoring status information 
a DC current of more than $3\,\mathrm{\mu A}$ (due to stars in the FoV) 
are also set to zero. The DC current values are 
provided together with the single pixel trigger rates in the normal data stream;
the information per pixel is updated online every 16 seconds.

We note that the exclusion of single faulty pixels does not impose strong inhomogeneity
problems in the reconstructed sky maps. Since defective pixels generally 
do not appear at the same sky positions for different telescopes, possible effects are averaged out.
Moreover, source candidates were generally observed offset from the
FoV center by 0\fdg5, resulting in further averaging of camera inhomogeneities
because of field rotation (due to the telescope alt-az mount).
On the other hand, if pixels are set to zero due to stars, these averaging effects do not apply;
noticeable distortions in the reconstructed sky maps of up to 20\% in the case of bright stars
can occur.
%%% Both is however not true for pixels which are set to zero due to stars; these can produce
%%% noticeable distortions in the FoV acceptance of up to 20\% in the case of bright stars.
This issue will be addressed in a forthcoming paper.

The exclusion of whole telescopes is deferred until the shower reconstruction is performed
on DST level. Here, 
single telescopes with technical problems are excluded. The most common reason for that is the
number of defective pixels. Telescopes
with generally more than 15 defective channels are excluded. 
In this case, usually at least one of the electronics modules -- each comprising 16 channels --
is broken. Since pixels which belong to the same module are clustered in the camera, 
the loss of one module leads to an inacceptable inhomogeneity in the respective camera.

The calibration procedure of the PMs and the respective electronics readout
which follows is mainly based on the
results of {\it laser runs} which are performed at the beginning of each observing night 
and occasionally repeated later on. For these calibration runs, 
nitrogen laser pulses are fed at each telescope into a scintillator, 
which in return illuminates the camera homogeneously with a spectrum which is
roughly similar to the Cherenkov light spectrum and time distribution of real showers.
Each {\it laser run} comprises 100 laser flashes
with a duration of a few nanoseconds and typical amplitudes of 80 to 100\,ph.e.
These values are far above the pedestal values, but below the level of 150\,ph.e.\ where
non-linear effects begin to play a role.

From the {\it laser runs}, the relative gains of the pixels are directly determined.
These results are used first of all
to adjust every few months the HV of all individual PMs.
The aim is to provide during data taking a trigger acceptance as homogeneous as
possible across the camera surfaces; this is needed to achieve good sensitivity for the 
compact shower images.
This HV flatfielding procedure leaves a 5\% trigger acceptance
variation across the camera FoVs,
due to sensitivity differences between the FADCs (which are calibrated) and discriminators
(which provide the trigger decision). 
Again, the field rotation and the averaging over many cameras, in conjunction with the usual
software thresholds of $40\,\mathrm{ph.e.}$ per image, led to the fact that no strong 
inhomogeneities in sky maps were induced by the trigger threshold differences.
%; however, this effect is negligible.} 

The gain differences between pixels in the FADC readout
which remain despite the HV adjustment are corrected during data
calibration, on the basis of the latest {\it laser run} result. The pixel amplitudes are 
multiplied with their respective correction factors,
which are close to unity with a RMS of about 5\%.

Also from the {\it laser runs},
the arrival time differences between pixels (a few ns) can be determined, since
the light from the laser arrives simultaneously at all PMs (Fig.\,\ref{F:lasercalib}, upper panel).
The signal propagation time depends on the HV value of a pixel, resulting in different PM transit
times (cf.\ Fig.\,\ref{F:lasercalib}, middle panel),
as well as on the hardware settings of the respective FADC channel
(cf.\ Fig.\,\ref{F:lasercalib}, lower panel).
These time differences are used in the pulse shape analysis which calculates the 
pixel amplitudes.
The following procedure is applied:
(a) in a first step, an event time is derived, using only pixels with a reasonably strong signal;
(b) in a second step, the respective 
expected time at which the signal is expected in each pixel is calculated from the event time.
In both cases, the relative propagation delay is corrected.
In this way, only the true event time is sampled also at channels with very weak pulses,
alleviating the influence of the night sky noise.
%The aim of this procedure is noise suppression.

The pixel timing constants were usually derived on a monthly basis. After HV adjustments or exchanges
of PMs or electronics modules, new tables were produced as well.
The timing calibration even permitted measurement of 
time profiles of air showers, with a duration of a few nanoseconds 
in the cameras \cite{HEGRAPaperOnShowerProfiles}.

This calibration scheme has worked robustly throughout the entire operational lifespan of the system.

\section{Telescope pointing}
\label{S:pointing}

During data taking, the tracking algorithm which steers the telescopes relies solely
on the position of the telescopes' axes, which are measured by optical shaft encoders.
%Each shaft encoder has a nominal resolution of $0\fdg022$. 
The steering implementation
is able to keep the telescopes on track within $\pm 1$ shaft encoder unit ($0\fdg022$).
However, inaccuracies of the telescopes' 
mechanical structures as well as non-linearities
of the shaft encoders lead to a typical online mispointing of $\approx 0\fdg05$, sometimes 
ranging up to $\approx 0\fdg3$. These pointing deviations are only depending on the 
orientations of the telescopes' axes, and are fully reproducible. Hence, they can be 
calibrated and corrected offline. 

Since the corrections are small compared to the camera size, online corrections are not required.
The coincidence rate loss and the system FoV inhomogeneities, caused by changing
overlaps of the camera FoVs at their borders, are also small.

\begin{figure}[t]
\vspace*{2.0mm}
\hc{\includegraphics[width=1.0\columnwidth]{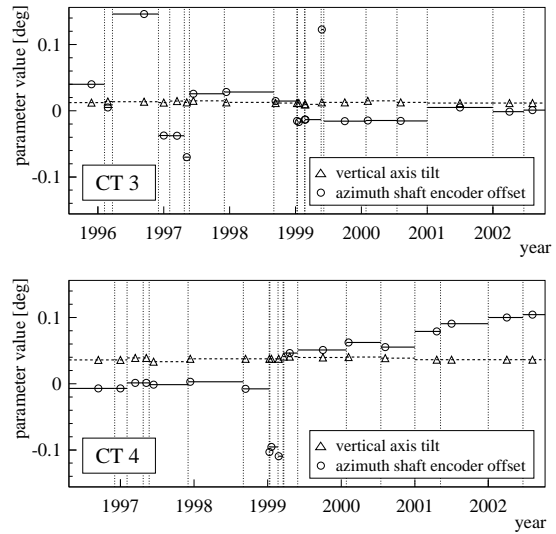}}
\caption{Evolution of two pointing model parameters for 
         %two telescopes,
         CT\,3 and CT\,4.
         The vertical dotted lines indicate the time spans during which one respective set 
	 of parameters was valid. Within errors, the vertical axis tilt has not changed
	 for any telescope (dashed line). 
	 Shaft encoder offsets have changed quite a few times due to repair work; 
	 additionally, for CT\,4 a small drift of the offset was observed after 1999.}
\label{F:pointingparameter}
\end{figure}

\subsection{Pointing corrections}
\label{SS:pointingcorrections}

The pointing of the system telescopes is corrected using an analytical
model of the telescopes' mechanical structure. 
This model allows one to compute the mispointing of each telescope for any given
elevation and azimuth angle. 
It parametrizes the bending of the masts and the telescope structure, zero offsets of
the optical axis and of the shaft encoders, tilts of the vertical and the horizontal axes,
and the first harmonic deviation from linearity of the shaft encoders.
For each telescope, 11 parameters 
%-- such as a slight tilt of the vertical axis or zero offsets of  the shaft encoders -- 
fully describe the pointing behaviour.  
The same model works
for all system telescopes, but different model parameters apply
for each of the five telescopes.
The systematic pointing error which remains after correction is about 25\arcsec\ in both
right ascension and declination;
this value was derived from the calibration procedure itself.

The procedure to derive the model parameters makes use of observing
selected bright stars as reference sources with the telescopes, by
performing so-called {\it point runs}. 
In one {\it point run}, a star spot is scanned across the central pixel of the camera;
13 horizontal scan lines cover a $0\fdg8 \times 0\fdg8$ window. The DC currents of the single pixels are
used to determine the actual pointing relative to the star.
One calibration procedure comprises a sample of stars distributed homogeneously across the sky.
More details are given in \cite{HEGRAPointing}. 

The initial pointing calibration was completed during the telescopes'
commissioning phase. Afterwards, the model
parameters needed to be redetermined only on special occasions,
such as a telescope mirror readjustment or a dismounting and remounting of the 
shaft encoders. 
Figure~\ref{F:pointingparameter} shows the temporal evolution of two exemplary parameters,
the vertical axis tilt and the zero offset of the azimuth shaft encoder, 
for two telescopes over their full lifetime. 
The vertical axis tilts of all telescopes are of the order of 0\fdg05,
in agreement with the accuracy which was envisaged during installation,
and have, within errors, not changed at any telescope.
The shaft encoder zero offsets, on the other hand, have changed a few times due to repair work.
For CT\,4, an additional small drift of the offset was observed (see Fig.\,\ref{F:pointingparameter}),
presumably starting after the last manual adjustment of this shaft encoder in 1999.

For the shower reconstruction, the mispointing is computed for each event, and 
the image center of gravities $\vec{x}_{tel}$ (see \S\,\ref{SSS:directionandcore}) are 
corrected in the camera plane. In principle,
the mispointing also leads to errors in the image orientations $\vartheta_{tel}$.
However, the actual mispointing values are small enough that the respective errors
of $\vartheta_{tel}$ are below the accuracy which can be achieved,
and the errors are therefore not corrected.

\subsection{The TeV center of gravity of the Crab}
\label{SS:pointingcrab}

\begin{figure*}[t]
\begin{center}
\hc{\includegraphics[width=0.6\textwidth]{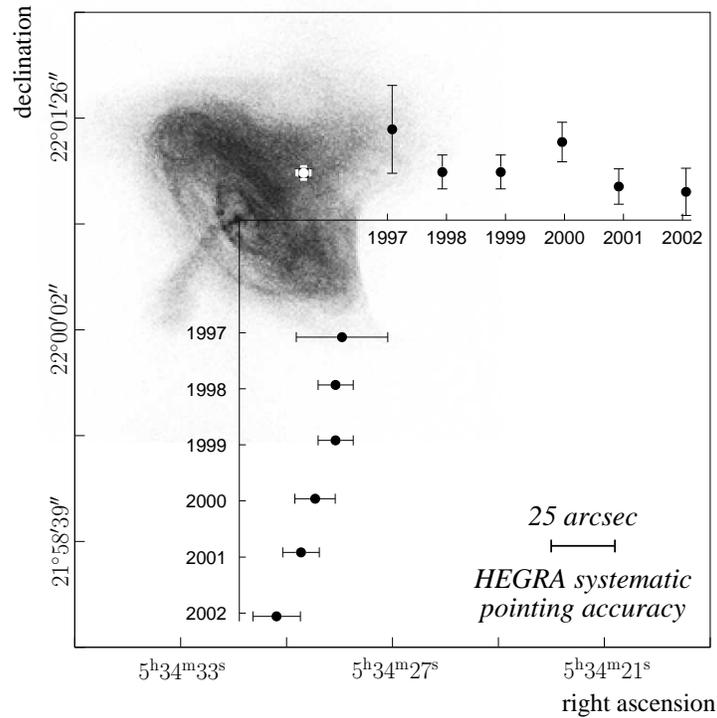}}
\caption{Positions of the X-ray and TeV $\gamma$-ray emission from the
Crab Nebula, shown in celestial coordinates. The grey-le\-ve\-led image shows
the recent Chandra X-ray image (courtesy of NASA/CXC/SAO);
the apparent position of the pulsar in the center of the image was placed 
at the year 2000-coordinates of the pulsar, as measured in the radio band.
For the TeV $\gamma$-ray emission, HEGRA data after a shape cut $msw < 1.1$ and 
below $30\degr$ zenith angle were used; data between 30\degr\ and 60\degr\ yield
within their reduced statistics consistent results. 
The white cross denotes the reconstructed center of the TeV emission with its 
statistical error for the entire data sample. Additionally, the deviation
from the expected position is shown as a function of the observation epoch,
both for declination (graph with horizontal axis) and for right ascension
(graph with vertical axis). 
%
%Each pair of a cross and a circle with error bars, connected with a dotted line,
%indicates the position of the X-ray pulsar
%and the fitted center of gravity of the TeV $\gamma$-ray emission, respectively, for
%five different observing seasons. The insets at the bottom show the position of the
%center of the TeV emission for all data (now corrected for precession) 
%at different zenith angle ranges.
}
\label{F:crabdevel}
\end{center}
\end{figure*}

The temporal stability and accuracy of the pointing calibration
throughout the entire operating lifetime of the telescope system
was verified for example by the examination of 
the center of gravity of the TeV emission of the Crab Nebula.
Figure~\ref{F:crabdevel} shows the sky region around the Crab Nebula
in celestial coordinates. For reference,
the Chandra X-ray image is shown. 
The apparent position of the Crab pulsar, being in
the center of the X-ray image, was placed in Fig.\,\ref{F:crabdevel}
at the year 2000-coordinates of the pulsar, as measured in the radio band.
The center of the TeV emission is determined by a fit of a two-dimensional
Gaussian distribution to all events surviving a $\gamma$-ray cut ($msw < 1.1$),
reconstructed in celestial coordinates. The typical angular resolution of the IACT system is
$0.1^\circ$, but with sufficiently high event statistics the center of
the TeV emission can be extracted with much better accuracy. In the case of Crab,
the statistical error of the reconstructed TeV center position
is 3.2 arcseconds in each coordinate for the total HEGRA data sample
(white cross in Fig.\,\ref{F:crabdevel}).

For a check of possible systematic errors in the telescope pointing,
the HEGRA data were split into six observation periods, from the year 1997 to the year 2002.
Figure~\ref{F:crabdevel} shows the deviation from the expected position
in declination (inset with horizontal axis) and right ascension (inset with vertical axis),
as a function of time. 
From the data one can conclude
that within the sum of statistical errors
($\approx 7\arcsec$ per year, $3\farcs2$ for the total sample)
and systematic errors ($25\arcsec$), the centers of the
X-ray and TeV $\gamma$-ray emission are well in agreement over the full lifetime of the experiment.
Although the statistics of the full data sample would permit, 
the systematic error does not allow to assign the center of TeV emission to either the
pulsar or the X-ray cloud.
However, the given pointing accuracy allowed to reliably search for extended TeV emission
from the Crab nebula. For that purpose, only $\gamma$-ray events with a 
predicted angular resolution of better than 0\fdg03 were used \cite{HofmannCrabSize}.

\section{The point spread function of the telescopes}
\label{S:pointspread}

The quality of the point spread function 
of each telescope is primarily determined by the alignment accuracy of the
individual mirror tiles
and by the slight deformations of the reflector itself at different elevations.
The mirror tiles are spherical glass mirrors of 60\,cm diameter; 
they are manually adjusted and fixed with screws. The adjustment was done as follows:
The telescope is pointed 14 degrees above horizon to a point light source
in approx. 900\,m distance. For each mirror, the appropriate distance to the mirror dish
and the alignment is found by minimizing the spot size in the focal point,
which lies in the focal plane corresponding to the light source.
After the alignment procedure, the camera is then shifted such that the front of the
funnel plate is focused to 8\,km distance.
Since the focal distance of the telescope is 4.92\,m, the correct focal plane is only 
displaced by 2.5\,cm from the focussing plane.

The HEGRA system telescopes are operated in two tracking modes:
normal and reverse mode. Reverse
mode means that the telescope is driven from (an arbitrarily defined) normal mode 
across the zenith in the opposite direction,
and thus operating upside down. 
While changing the azimuth direction does not noticeably influence the mount structure, 
changing the zenith angle puts different stress on the mount.
Since the mirror alignment was usually done
in normal mode close to the horizon, it must be ensured that deformations of the telescope
mounts do not significantly change the {\it psf} of the reflector across the whole observing
range. This was part of the mount specifications, 
and regular observations up to 60\degr\ zenith angle were performed. 
%%% Optimum zenith angles for data taking are below 45\degr, 
%%% regular observations up to 60\degr\ were
%%% however possible and sometimes
%%% performed.

\subsection{Direct monitoring of the point spread function}
\label{SS:monitoringpointspread}

\begin{figure*}[t]
\begin{center}
\hc{\includegraphics[width=0.85\textwidth]{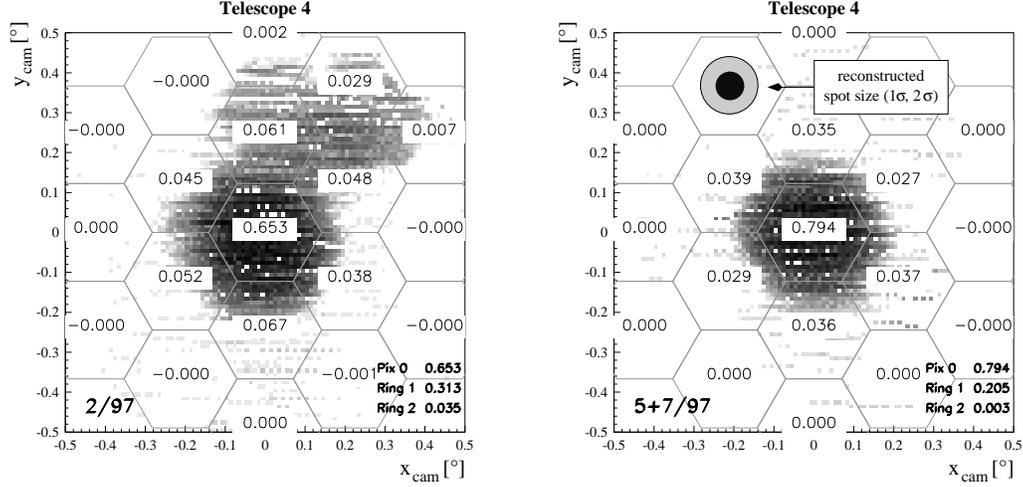}}
\caption{Indirect measurement of the point spread function of an individual telescope;
         both images were obtained summing up different sets of {\it point runs}.
         Encoded in grey levels is the average PM current of the central pixel as a function
	 of the displacement of the camera center to the respective star position. The currents are a
         measure of the convolution of the {\it psf} with the hexagonal surface function.
         The pixel size is indicated by the hexagonal camera structure,
         the reflector spot size itself is smaller than one pixel as indicated by the reconstructed spot size.
         The numbers in the pixel centers quote the integrated current which falls into the respective
	 surface area, normalized to the total distribution;
%	 relative amount of light which falls into the respective pixel,
         by definition, this is equivalent to the percentage of light which hits the respective
	 pixel if a light beam was homogeneously distributed across the surface of the central pixel.
         Hence, for a well-adjusted mirror (right panel), on average 80\% of the light of a point source
	 (at random position in a pixel)
         actually hits the respective pixel (''Pix 0''),
%	 if the focus is homogeneously distributed across its surface; 
	 while 20\% is spread across the adjacent 6 pixels (''Ring 1''). 
	 The presence of misadjusted mirrors (as
	 for example 5 mirrors on the left panel) can easily be detected in this representation.
	  }
\label{F:spotshape}
\end{center}
\end{figure*}

\begin{figure}[t]
%\begin{center}
\hc{\includegraphics[width=1.0\columnwidth]{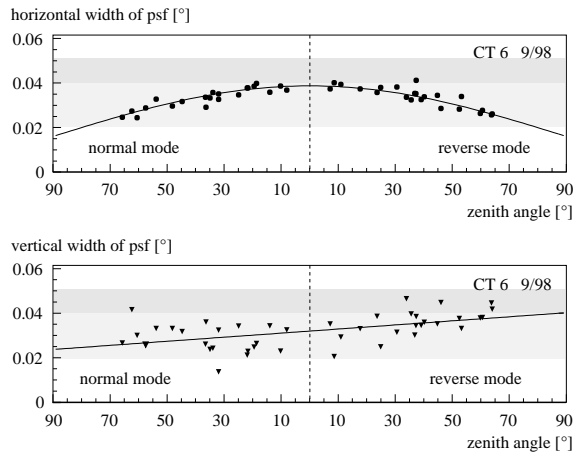}}
\caption{The width of the reflector spot size as a function of the zenith angle in normal and reverse mode,
         here for CT\,6 in fall 1998.
         The spot size is characterized by the width of a two-dimensional Gaussian
         with different values for the horizontal axis (upper panel) and the vertical axis (lower panel).
	 The accuracy of a single measurement can be estimated from the scatter; as expected from the 
	 way of measurement, it is much better in horizontal than in vertical direction. The lines show
	 fits of empirical representations for the typical dish behaviour to the data points.
	 The light grey band denotes the range in which experience showed that regular operation was guaranteed.
         A distortion of the mirror alignment shifts the {\it psf} to higher values;
         if the {\it psf} entered the dark grey band, a realignment of the mirrors was recommended.}
\label{F:reflectorbend}
%\end{center}
\end{figure}

The {\it psf} can be characterized by its on-axis width; 
the additional off-axis aberrations are purely of geometrical origin 
and are not influenced by changes of the mirror adjustment quality or reflector bending.
The dependence of the {\it psf} on the distance from the optical axis was measured
during the commissioning of the telescopes, by scanning stars across the entire cameras.
The parametrisation thus derived was compared via Monte Carlo simulations with full ray-tracing
simulations; the simulated image parameter distributions were fully compatible \cite{MarcThesis}.

In general, the imaging quality of all telescopes was stable throughout the years,
also with the help of mirror
readjustments which were performed roughly once every year.
Some problems occurred due to icing of the telescopes which has lead to distortions of
the mirror tiles' alignment a few times; this could be cured by subsequent mirror readjustments.

The deterioration of the {\it psf} was monitored with
{\it point runs}.
The on-axis {\it psf} can be measured by standard {\it point runs}
which are used for pointing calibrations (see \S\,\ref{S:pointing}),
where a star is scanned across the central pixel.
Whereas a single {\it point run} is not sensitive enough to trace single
misadjusted mirrors, several runs can be combined to produce images such as shown in 
Fig.\,\ref{F:spotshape}. Here, the {\it psf} convoluted with the hexagonal surface of the central pixel 
is represented in grey levels; differences of the {\it psf}
at different elevation angles -- due to the reflector bending -- are averaged out.
For a well-adjusted reflector, on average 20\% of all photons
which should hit the central pixel are -- due to the finite spot size -- imaged into adjacent pixels 
(Fig.\,\ref{F:spotshape}, right panel). In this representation, single or several misadjusted mirrors can
easily be traced, as shown for example in the left panel of Fig.\,\ref{F:spotshape}; in this particular case,
five mirrors were completely misfocused.

From a single {\it point run}, only the width of an assumed two-dimensional 
Gaussian {\it psf} can be determined,
for the vertical and the horizontal axis of the spot. To derive these numbers, the Gaussian is convoluted
with the hexagonal pixel surface area and then fitted to the measured DC current in the central pixel.
The results of such an analysis,
derived from all {\it point runs} of one complete pointing calibration campaign, are shown in 
Fig.\,\ref{F:reflectorbend} as a function of the zenith angle.
Due to the way {\it point runs} are performed --
horizontal scan lines of a star spot across the central pixel of the camera -- the resolution
in horizontal direction (upper panel of Fig.\,\ref{F:reflectorbend}) is much better than in vertical
direction (lower panel).

The best -- i.e.\ smallest -- {\it psf} is obtained close to 90\degr~zenith angle in normal mode,
as expected since mirror adjustments are done near this position. Yet, 
as already mentioned above, the mirror dishes were designed to
be stiff enough to provide an adequate {\it psf} over the full zenith angle range.
The measurements showed that the telescopes CT\,3 to CT\,6 have a {\it psf} with
a Gaussian width of $\sigma_{\mathrm{psf}} = 0\fdg03..0\fdg04$ per axis in the 
relevant zenith angle range \cite{GerdThesis}.
CT\,2 however has a weaker reflector structure; it was designed as a prototype for the
forthcoming system telescopes, and was integrated last into the system in summer 1998,
after having run as a standalone telescope for several years before.
Then, after initial measurements, it was decided to align the mirrors in vertical telescope
position to achieve best alignment quality near typical -- i.e.\ small -- observation zenith angles.
Nevertheless, the {\it psf} of CT\,2 has always been inferior to the other telescopes, 
and made a special data treatment necessary (see \S\,\ref{SS:shapeexpectation}).

\subsection{Shape parameter expectation values}
\label{SS:shapeexpectation}

The quality of the {\it psf} is an important parameter for the 
shape of the images that are recorded from air shower events.
Detector simulations have been applied to simulated showers in order
to predict expectation values for image shape parameters
for $\gamma$-rays (and also for charged cosmic rays) \cite{HEGRAPerformance99}.
The simulated values were extensively compared to real $\gamma$-ray events obtained from 
the Crab nebula, and especially from Mkn\,501 during its great outburst in summer 1997 
which provided a quasi background-free $\gamma$-ray sample.
Simulations and experimental values were found in general to be in excellent agreement 
\cite{Mrk501PaperI,HEGRAPerformance99}.

Usually, the information about the shape of a shower event is combined into the single parameter
{\it mean scaled width} ({\it msw}, see \S\,\ref{S:dataprocessing}). 
Figure~\ref{F:meanscaledwidth} shows the distribution of this parameter
for $\gamma$-rays and background events,
obtained from Mkn\,501 (background-subtracted) and from off-source background measurements,
respectively.
The distribution for $\gamma$-rays should, by definition, be centered at $msw_0 = 1.0$. The shape is roughly Gaussian,
with a width of $\sigma = 0.1$; a more accurate description can be achieved by
adding a second Gaussian
with 20\% of the amplitude and a width of $\sigma_2 \approx 0.15$.
The expected distribution for $\gamma$-ray events is needed (a) to predict the acceptance $\kappa_{\gamma,msw}$ 
of the cut which is applied to reject cosmic ray events (usually $msw < 1.1$ or $msw < 1.2$), and (b) in combination
with the background distribution to calculate optimized cuts for the signal search.

Precise investigations showed that the value of $msw_0$ for real $\gamma$-ray data without any corrections
actually was 1.033. 
Hence, in order to achieve the highest possible agreement between the acceptances predicted from simulations 
and obtained from real data, the expectation values for $\gamma$-rays were multiplied by a 
correction factor $c_{msw}=1.033$;
the distributions shown in Fig.\,\ref{F:meanscaledwidth} already include this correction.
Apart from that, the shape expectation values have remained constant throughout the
lifetime of the telescope system and required no further treatment,
with the exception of (a) a few periods where misfocused mirrors
led to measurable distortions of the {\it psf} (see \S\,\ref{SS:monitoringpointspread}), and
(b) a special treatment of CT\,2.

\begin{figure}[t]
%\begin{center}
\hc{\includegraphics[width=1.0\columnwidth]{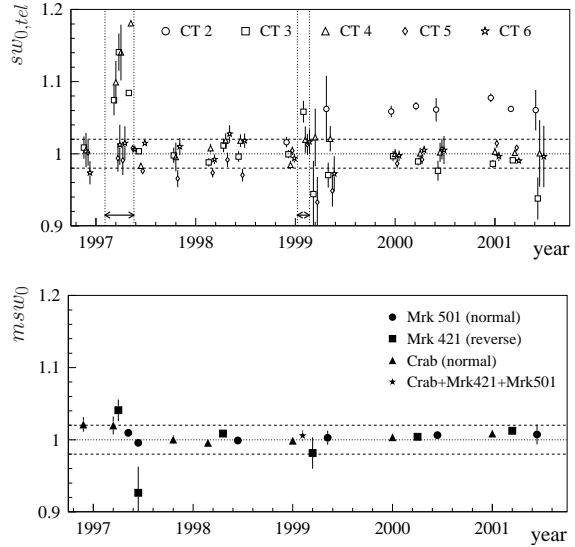}}
\caption{Upper panel: Time evolution of the mean values $sw_{0,tel}$ of the {\it scaled
         width} distributions. 
%        of all system telescopes. 
         The values were derived from
	 $\gamma$-ray data from Crab, Mkn\,501 and Mkn\,421 (cf.\ lower panel). The vertical
	 dotted lines indicate two periods with strongly misadjusted mirrors: CT\,3
	 and CT\,4 in the first, all telescopes (mainly CT\,3) in the second case.
	 Lower panel: The mean value $msw_{0}$ of the {\it mean scaled
         width} distribution, after application of correction factors $c_{msw,tel}$
         to individual telescopes, essentially remained constant throughout the years.
         All values in this figure were computed using Monte Carlo expectation values which
	 already include the correction factor $c_{msw}=1.033$ (see text for details).
	 }
\label{F:shapedevel}
%\end{center}
\end{figure}

Distortions of the {\it psf} (as e.g.\ shown in Fig.\,\ref{F:spotshape}, left panel) 
lead to a loss of events that would otherwise correctly be identified as $\gamma$-rays,
since $\gamma$-rays are expected to produce smaller images than charged cosmic rays.
Figure \ref{F:shapedevel} shows in the upper panel the mean value $sw_{0,tel}$ 
of the {\it scaled width} distribution for individual telescopes 
(i.e.\ before averaging over all telescopes, see \S\,\ref{S:dataprocessing}). 
The values were obtained from $\gamma$-ray events from the Crab nebula, Mkn\,501, and Mkn\,421,
in intervals of yearly observing seasons with the exception of periods with known
distortions of the {\it psf} at some or all telescopes. 
We note that with the exception of short periods and of CT\,2, the scaled widths average within
$\pm 2\%$ of 1.
Two periods with the strongest temporary deviations from the expected value 1.0 
are indicated in the Figure by vertical dotted lines.
Besides that, CT\,2 permanently shows mismatching values for $sw_{0,tel}$
for most of its lifetime.

System data or data from individual telescopes 
which were affected by distorted point spread functions were either discarded, or the respective
expectation values were corrected by additional factors $c_{msw,tel}$ (e.g.\ \cite{Mrk501PaperI}).
The values for $msw_0$ which include all corrections 
are plotted in Fig.\,\ref{F:shapedevel}, lower panel.
As can be seen, the expectation values $msw_0$ for $\gamma$-rays after corrections
were essentially stable at 1.0
throughout the years. Hence, a constant $\gamma$-ray acceptance $\kappa_{\gamma,msw}$ 
for the usual cuts was provided.
However, when including telescopes with a poorer {\it psf}, an
increased background level due to the inferior separation power of the imaging
has to be accepted.

\section{Relative sensitivity monitoring}
\label{S:relative}

The overall detector sensitivity needs to be monitored continuously in
order to perform the energy calibration of the instrument as well as
to calculate the actual collection area for $\gamma$-rays.
In this section, we discuss the calibration relative to the reference
periods 62-69 (second half of 1997);
\S\,\ref{S:absolute} addresses the absolute sensitivity calibration.

\subsection{The photon conversion efficiency: optical and electronic gain}
\label{SS:conversion}

\begin{figure*}[t]
\begin{center}
\vskip 1.5pc
\hc{\includegraphics[width=0.9\textwidth]{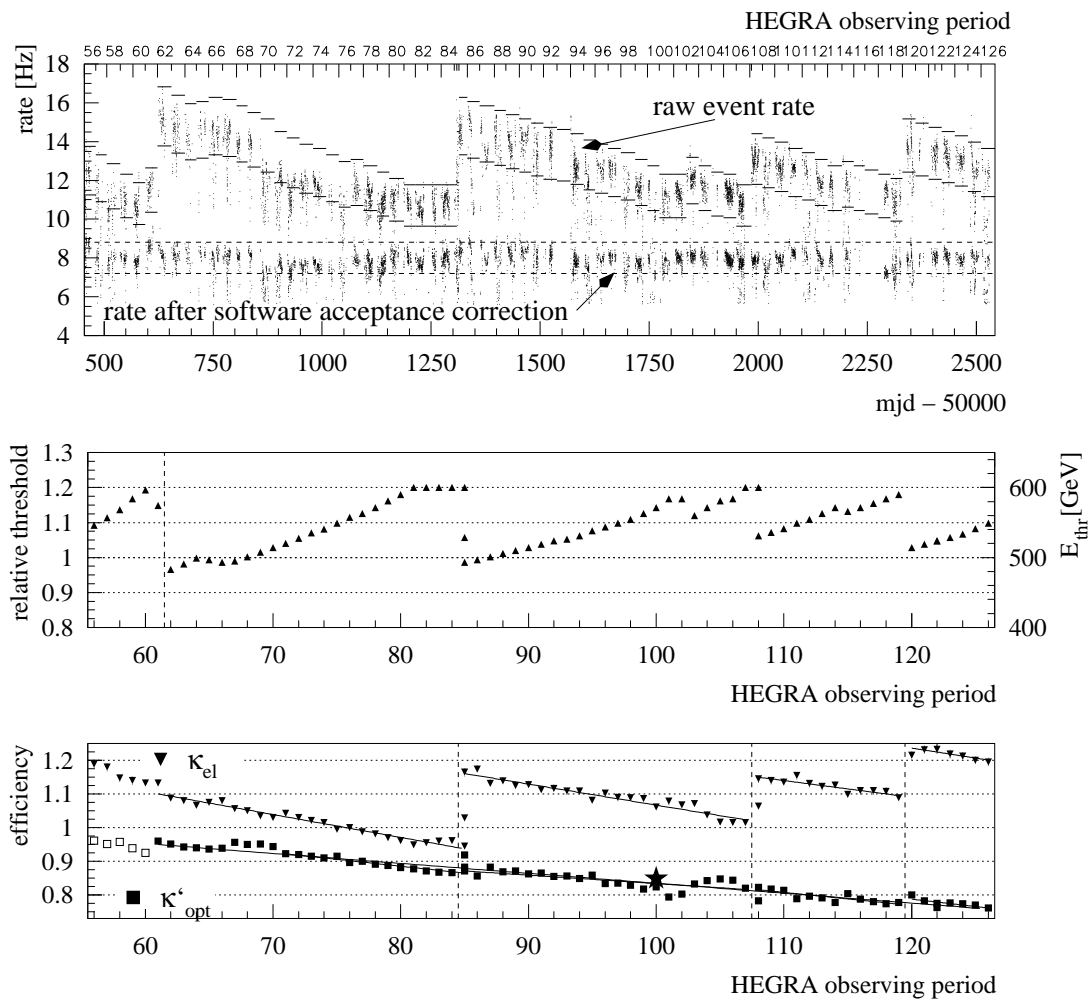}}
\vskip 1.5pc
\caption{Upper panel: System trigger rate (converted to a 4-telescope rate), and event rate after 
a software acceptance cut.
Each dot represents one data run of typically 20\,min; 
runs up to 30\degr\ zenith angle were used.
The lines indicate the $\pm 10\%$ range of the respective expectation values.
Middle panel: The deduced change of the energy threshold relative to periods 62-69. 
For the right energy scale, 500\,GeV was assumed to be the nominal threshold.
Lower panel: Photoelectron-to-FADC-count conversion factor
$\kappa_{\mathrm{el}}$, averaged over all telescopes, and relative optical efficiency
$\kappa'_{\mathrm{opt}}$ of the system. 
The vertical dashed lines denote global HV increases of all cameras.
The star indicates the result of direct mirror reflectivity measurements.
The plot ranges from January 1997 to September 2002. For details see text.}

\label{F:ratedevel}
\end{center}
\end{figure*}

The sensitivity of a Cherenkov telescope is mainly determined by the total
mirror area, the reflectivity of the mirrors, the quantum efficiency
and gain of the PMs, and finally by the trigger threshold. 
We express the efficiency by which the Cherenkov photons are converted into 
signal amplitudes with two values: 
\\
$\bullet$
The optical efficiency $\kappa_{\mathrm{opt}}$
comprises the conversion of the number of Cherenkov photons which hit the reflector area under 
a certain solid angle
to the number of photoelectrons released by the PM cathode of the corresponding 
pixel\footnote{Ideal imaging is assumed.}.
%; this is however not important for the following}.
In this section, we only deal with a relative factor $\kappa'_{\mathrm{opt}}$ close to 1,
which describes the efficiency change relative to some nominal value.
The values for $\kappa'_{\mathrm{opt}}$ are only derived per period on average for all telescopes 
in the standard calibration procedure; the method is based on the trigger rate development of the system 
as explained in the following subsection (\S\,\ref{SS:rateandoptical}).
For differences between telescopes see the results in \S\,\ref{SS:muonruns}.
\\
$\bullet$
The PM plus electronic amplification in terms of a photoelectron-to-FADC-count conversion factor
$\kappa_{\mathrm{el},tel}$ can be determined for each individual telescope.
The values can be derived from the {\it laser runs}
discussed in \S\,\ref{S:camera} \cite{MarkusThesis}.
The method uses the fact that the variations of the measured pixel 
amplitudes from laser shot to laser shot -- after correction of the laser intensity fluctuations --
are a measure for the average number of released photoelectrons;
other contributions to the width of the signal spectrum can be calibrated in the laboratory.
%the method makes use of additional information obtained from the {\it laser runs}
%discussed in \S\,\ref{S:camera} \cite{MarkusThesis}.
These electronic conversion factors can be measured absolutely, and were set
close to one FADC count/ph.e.\ during the commissioning phase of the telescopes,
by adjusting electronic gains and HV values of the PMs.
The values for $\kappa_{\mathrm{el},tel}$ 
are determined for the system calibration on a period by period basis.

Since all telescopes show nearly the same time dependence of $\kappa_{\mathrm{el},tel}$
due to common aging processes,
an average value $\kappa_{\mathrm{el}}$ can be reasonably calculated; 
it is determined according to
\begin{equation}
   \label{G:averagekappael}
   \kappa_{\mathrm{el}} = \frac{1}{N_{tel}+1}~\left(\kappa_{\mathrm{el,CT\,3}} +
   \sum_{tel}{\kappa_{\mathrm{el},tel}}\right)
\end{equation}
The sum runs over all $N_{tel}$ telescopes included in the system; 
the central telescope CT\,3 is assigned double weight,
which reflects (very roughly)
that it is overabundant in the events because of its position at the system center.
%For $\kappa'_{\mathrm{opt}}$, the validity of the averaging over all telescopes can only be assumed.

The values for $\kappa_{\mathrm{el}}$ and
$\kappa'_{\mathrm{opt}}$ are both shown in Fig.\,\ref{F:ratedevel}, lower panel,
for the whole lifetime of the telescope system.
With time, $\kappa_{\mathrm{el}}$ degrades continuously, presumably due to aging of the PM dynodes.
The rise of $\kappa_{\mathrm{el}}$ at the positions of the vertical dashed lines reflect
global increases of the HV in all cameras; these were performed to compensate for the previous gain
losses. For $\kappa'_{\mathrm{opt}}$, a continuous slow degradation is observed.
From the variations of the data between adjacent periods, we can conclude that the relative sensitivity monitoring
has an accuracy of a few percent.

\begin{table}[t]
\begin{center}
\begin{tabular}{lll} \\ \hline \hline
  \multicolumn{2}{l}{signal}                           & unit                \\ 
  & $\times$ conversion factor                         & $\approx$ value     \\  \hline
  \multicolumn{2}{l}{photon flux (300-600\,nm)}        & [ph.]                   \\
  & $ \times~\kappa^{*}_{\mathrm{opt}} $               & $\approx 0.12\,\mathrm{ph.e.}^{*}/\mathrm{ph} $ \\
  \multicolumn{2}{l}{nom. photoelectron counts}        & [$\mathrm{ph.e.}^{*}$]  \\
  & $ \times~\kappa'_{\mathrm{opt}} $                  & $\approx 1\,\mathrm{ph.e.}/\mathrm{ph.e.}^{*} $ \\
  \multicolumn{2}{l}{act. photoelectron counts}        & [ph.e.]                 \\
  & $ \times~\kappa_{\mathrm{el}} $                    & $\approx 1\,\mathrm{FADC}/\mathrm{ph.e.} $ \\
  \multicolumn{2}{l}{FADC counts}                      & [FADC]                  \\ \hline \hline
  & $ \kappa_{\mathrm{opt}} = \kappa^{*}_{\mathrm{opt}} \cdot \kappa'_{\mathrm{opt}} $ 
                                                       & $\approx 0.12\,\mathrm{ph.e.}/\mathrm{ph} $ \\
  & $ \kappa'_{\mathrm{tot}} = \kappa_{\mathrm{el}} \cdot \kappa'_{\mathrm{opt}} $ 
                                                       & $\approx 1\,\mathrm{FADC}/\mathrm{ph.e.}^{*} $ \\
  & $ \kappa_{\mathrm{tot}} = \kappa_{\mathrm{opt}} \cdot \kappa_{\mathrm{el}} $ 
                                                       & $\approx 0.12\,\mathrm{FADC}/\mathrm{ph} $ \\ \hline \hline \\
\end{tabular}
\end{center}
\caption{The table sketches how the signal is converted, together with the corresponding
         conversion factors and their typical values.
	 In the reconstruction, one starts from the bottom beginning
	 with the FADC measurement, and computes the number of photons to determine the shower energy.}
\label{T:conversionfactors}
\end{table}

\subsection{The system event rate and the relative optical efficiency}
\label{SS:rateandoptical}

As already mentioned, the optical efficiency $\kappa'_{\mathrm{opt}}$ itself was not measured directly.
Instead,
the cosmic-ray induced event rate is a good measure for the overall detector threshold and sensitivity;
the system practically does not trigger on noise events or local muons \cite{HEGRATriggerPaper}.
The event rate $R$ of the system is shown in Fig.\,\ref{F:ratedevel}, upper panel;
for further processing, the rate was already normalized to a 4-telescope setup rate, using the
empirical relation
$R_{3 \mathrm{tel}}:R_{4 \mathrm{tel}} : R_{5 \mathrm{tel}} = 0.8 : 1 : 1.15$.
%In order to unveil the sensitivity out of the trigger rate,
To determine the detector sensitivity from the event rate,
this number needs to be corrected for weather influences,
and the actual dead time of the camera/trigger electronics must be taken into
account. After these (small) corrections, any change of the trigger event rate $R$ -- 
relative to the nominal rate $R_{0}=15\,\mathrm{Hz}$ -- can be directly
used to recalculate the energy threshold $E_{\mathrm{thr}}$, 
relative to the nominal energy threshold of $E_{\mathrm{thr,0}}=500\,\mathrm{GeV}$: 
\begin{eqnarray}
  R \propto F_{\mathrm{CR}}(E>E_{\mathrm{thr}}) \propto E_{\mathrm{thr}}^{-1.7} \\
  \label{G:threshfromrates}
  E_{\mathrm{thr}}/E_{\mathrm{thr,0}} = (R_0/R)^{0.58}.
\end{eqnarray}

We note that early measurements of the system trigger rate with different
hardware threshold settings resulted in a power-law behaviour with
a spectral index 
%an effective cosmic ray spectral index
of $\alpha \approx -1.35$ for the standard system trigger settings
($ \ge 2$ neighboured triggered pixels in a camera, $\ge 2$ telescopes with camera trigger)
\cite{HEGRATriggerPaper}.
Monte Carlo simulations showed as well trigger rate dependencies with spectral indices above the
value for CRs, albeit on average 
%with indices that were
by $\Delta \alpha \approx 0.15$ 
steeper than the experimental data \cite{HEGRAPerformance99}.

Using the value of $\alpha = -1.35$ 
for the threshold estimate in Eq.\,\ref{G:threshfromrates} would lead to slight
discontinuities in the
derived development of $\kappa'_{\mathrm{opt}}$ 
(cf.\ Fig.\,\ref{F:ratedevel}, lower panel). The value of
$\alpha=-1.7$ (or even a bit lower) is preferred by this investigation.
To achieve consistency,
a threshold dependence $E_{\mathrm{thr}} \propto \kappa^{-1.25}$ (cf.\ Eq.\,\ref{G:threshfromkappa})
would be required. 
Given that the energy threshold does not correspond to a step function in the trigger response,
one can expect slightly different effective spectral indices for the different
ways of changing the threshold: on the one hand by setting different hardware threshold levels,
on the other hand by changing the overall gain of the detector.
Whether this is the reason for the slight discrepancy in the spectral indices is unknown.
However, the results of the present analysis are only weakly influenced by the exact value of the
effective CR spectral index.

The nominal threshold $E_{\mathrm{thr,0}}$ was 
calculated by simulations for the reference detector conditions 
$\kappa_{\mathrm{el}} = 1\,\mathrm{FADC}/\mathrm{ph.e.}$ and 
$\kappa'_{\mathrm{opt}} = 1\,\mathrm{ph.e.}/\mathrm{ph.e.}^{*}$ \cite{HEGRAPerformance99}.
The time development of the energy threshold is plotted in 
Fig.\,\ref{F:ratedevel} middle panel;
we can conclude that the energy threshold of the telescope system 
has not exceeded 600\,GeV throughout the entire lifetime of the experiment.

The detector sensitivity change can be explained to a large extent with the 
change of $\kappa_{\mathrm{el}}$. 
The remaining correction factor which is needed to understand the full
change of the sensitivity is assigned to the optics (mainly aging of the mirror surfaces);
thereby $\kappa'_{\mathrm{opt}}$ is determined \cite{GerdThesis}.
By definition, the time interval comprising periods 62-69 has reference 
detector conditions, $\kappa'_{\mathrm{opt}}$ is calculated such that
%%% \begin{equation}
  $\kappa'_{\mathrm{tot,62-69}} = 1\,\mathrm{FADC}/\mathrm{ph.e.}^{*}$
%%% \end{equation}
on average during this period, with $\kappa'_{\mathrm{tot}}$ defined as 
\begin{equation}
  \label{G:kappatotprime}
  \kappa'_{\mathrm{tot}} = \kappa_{\mathrm{el}} \cdot \kappa'_{\mathrm{opt}}.
\end{equation}
We use the simple estimate
\begin{equation}
  \label{G:threshfromkappa}
  E_{\mathrm{thr}}/E_{\mathrm{thr,0}} = (\kappa'_{\mathrm{tot}}[\mathrm{FADC}/\mathrm{ph.e.}^{*}])^{-1}
%  E_{\mathrm{thr}}/E_{\mathrm{thr,0}} = (\kappa_{\mathrm{el}}\mathrm{[FADC/ph.e.]} \cdot \kappa'_{\mathrm{opt}})^{-1}
\end{equation}
to translate the energy threshold change into an efficiency change.
%\footnote{direct measurements of the
%rate with different hardware threshold settings showed $E_{\mathrm{thr}} \propto \kappa^{-0.8}$, close
%to the functionality which was assumed here}.

For the detector calibration, 
$\kappa'_{\mathrm{opt}}$ is computed on a monthly basis from the measured rate $R$ and the measured
electronic conversion factors $\kappa_{\mathrm{el},tel}$
by combining
equations \ref{G:averagekappael}, \ref{G:threshfromrates}, \ref{G:kappatotprime} and \ref{G:threshfromkappa}:
\begin{eqnarray}
  \kappa'_{\mathrm{opt}} &=&  \left(\frac{R}{R_0}\right)^{0.58} \hspace{1ex} \frac{\mathrm{FADC}}{\mathrm{ph.e.}^{*}}  \\
                         &/&  \left\{ \frac{1}{N_{tel}+1}~\left(\kappa_{\mathrm{el,CT\,3}} +
			      \sum_{tel}{\kappa_{\mathrm{el},tel}}\right) \right\}. \nonumber
  \label{G:derivekappaopt}
\end{eqnarray}
The absolute scale of $\kappa'_{\mathrm{opt}}$ is 
%thereby chosen in 
arbitrary.
%units.
However, an extrapolation back to the start of the telescope system (roughly period 47) yields by chance 1;
$\kappa'_{\mathrm{opt}}$ can thus be interpreted as the combined quality degradation of 
all optical components (mirrors, plexiglass cover of the camera, funnel plate, and
quantum efficiency of the PMs) from the initial value.

Other values which influence the trigger rate remained unchanged:
the hardware threshold of the telescope system has remained unaltered
after its final adjustment to 8\,mV in May 1997;
also the total mirror area and the mirror alignment remained basically constant.

\subsection{Energy and spectral acceptance calibration}
\label{SS:energycalib}

The calibration values $\kappa_{\mathrm{el},tel}$ and $\kappa'_{\mathrm{opt}}$ are determined on a monthly basis 
and used for the energy calibration of the system data in two ways:
\\
$\bullet$
The image amplitudes are derived according to 
\begin{eqnarray}
%  c^{*}_{pix}   &=& \left(\kappa_{\mathrm{el},tel}\right)^{-1} c_{pix}                       \\
  amp_{tel}     &=& \sum_{pix} \left(\kappa_{\mathrm{el},tel}\right)^{-1} c_{pix}     \label{Eq:ampsumpix} \\
  amp^{*}_{tel} &=& \left(\kappa'_{\mathrm{opt}}\right)^{-1} amp_{tel} 
\end{eqnarray}
where $c_{pix}$ are the raw pixel amplitudes
and the sum runs over all values of $\left(\kappa_{\mathrm{el},tel}\right)^{-1} c_{pix}$
%$c^{*}_{pix}$ 
above the tailcut threshold 
(see \S\,\ref{SS:imageprocessing}).  
The rescaled values $amp^{*}$ 
are used in the event energy calculation (see \S\,\ref{SSS:energyreco}).
\\
$\bullet$
As the efficiency change also leads to a trigger acceptance variation, the values are used 
to rescale the energy dependence of the trigger probability $P_{\gamma}(E,r)$,
simply by rescaling the effective area $A_{\mathrm{eff}}(E)$ (see \S\,\ref{SSS:spectraleval})
according to 
\begin{equation}
  A^{*}_{\mathrm{eff}}(E) = A_{\mathrm{eff}}(\kappa'_\mathrm{{tot}} \cdot E),
\end{equation}
where $A_{\mathrm{eff}}$ is the effective area as derived by Monte Carlo simulations (see \S\,\ref{SSS:spectraleval}).

% and hence the effective area $A_{\mathrm{eff}}(E)$ (see \S\,\ref{SSS:spectraleval}). 

This procedure is valid provided that 
(a) the trigger acceptance functionality $P_{\gamma}(E,r)$ just scales
linearly within the experimentally given variations of $\kappa'_{\mathrm{tot}}$, and that 
(b) nonlinear effects (such as the tail cut threshold which is for technical reasons
computed on the ph.e.-scale, see Eq.\,\ref{Eq:ampsumpix}) do not play a significant role
for the given range of $\kappa'_{\mathrm{opt}}$.

%the first order (linear) scaling factor
%$\kappa_{\mathrm{el}}\cdot\kappa'_{\mathrm{opt}}$ can be passed through from the amplitudes to the
%effective area.

\subsection{Checks on the relative acceptance calibration}
\label{SS:relativecheck}

\begin{figure}[t]
%\begin{center}
%\includegraphics[width=6cm]{crab_rate}
\hc{\includegraphics[width=1.0\columnwidth]{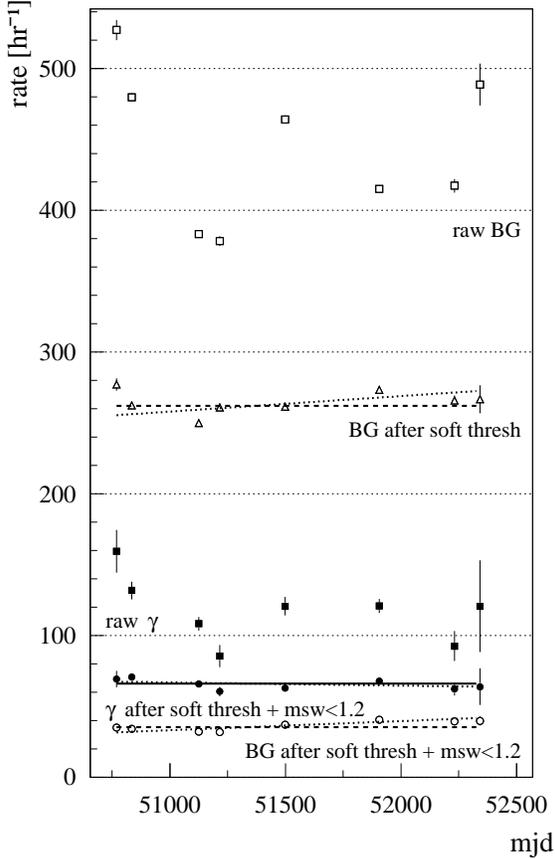}}
\caption{Event rates in the Crab field of view ($< 20\degr$ zenith angle),
         binned in time intervals with similar
         detector performance.
         Open symbols denote background (BG) rates, normalized to a circle with
         radius $\theta=0\fdg22$. Filled symbols represent the $\gamma$-rate
         from the position of the Crab,
         with an opening angle of $\theta=0\fdg22$;
         these rates are background-subtracted.
         Squares denote the raw rates without any software cut. Triangles are
         rates after a software acceptance threshold, as discussed in the text,
         circles additionally include a loose cut of $msw<1.2$.
%         The $\gamma$-rate after software threshold without $msw$-cut is 5\% higher than
%        with $msw$-cut (however with a larger statistical error due to the much larger background
%        level); the points were omitted for clarity reasons.
         To obtain comparable results over the years, CT\,2 was excluded, and
         only data with 4 active telescopes were used.
         }
\label{F:crabrate}
%\end{center}
\end{figure}

\begin{figure}[t]
%\begin{center}
\hc{\includegraphics[width=1.0\columnwidth]{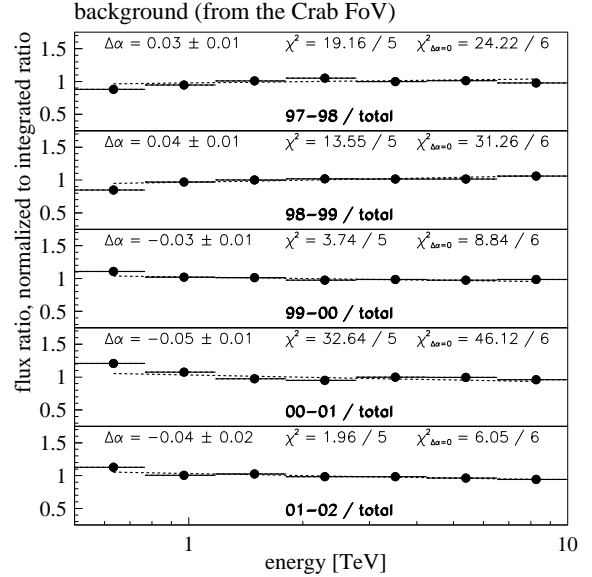}}
\caption{Ratios of background energy spectra 
         obtained from the Crab FoV. Only 4-telescope data (excluding CT\,2) up to 25\degr~zenith angle were used, 
	 a software threshold as discussed in the text was applied.
         The data were split into 5 observation seasons 97/98 to 01/02. 
	 For each event, the energy was reconstructed under the $\gamma$-ray hypothesis.
         A raw energy spectrum 
	 (i.e. without spectral acceptance correction) was filled for each time interval,
	 and normalized to the total event number to correct for different observation times.
         Each of the histograms was then binwise divided by the similarly computed histogram of the total data sample.
%         was  and compared to the total data sample. 
%         The data were normalized to the respective integrated ratios to correct for different
%	 observation times.
	 $\Delta\alpha$ denotes
	 the results of power law fits $\propto E^{\Delta\alpha}$ to the ratio histograms,
	 shown by the dashed lines. Statistical errors of the data are below the point size. 
         }
\label{F:crabspeccr}
%\end{center}
\end{figure}

\begin{figure}[t]
%\begin{center}
\hc{\includegraphics[width=1.0\columnwidth]{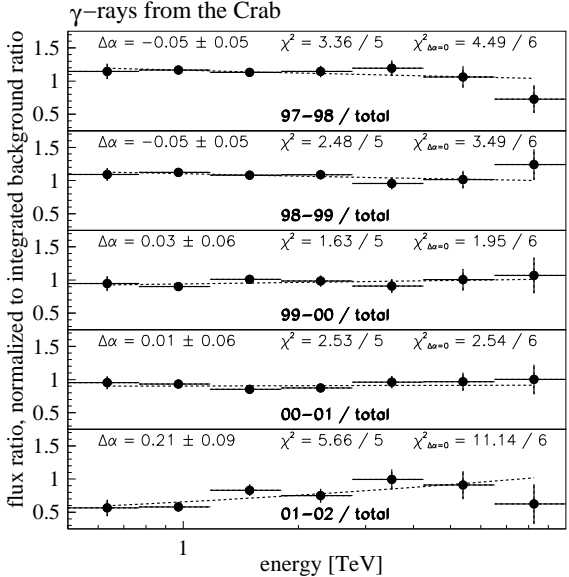}}
\caption{Ratios of $\gamma$-ray energy spectra 
         obtained from the Crab nebula. 
	 See the caption of Fig.\,\ref{F:crabspeccr} for how the histograms were obtained. 
	 Here, events from the direction of the Crab nebula were used, after applying a shape cut
	 $msw <1.1$ to enhance the $\gamma$-ray signal, 
	 and using background control regions to subtract the remaining CR background.}
\label{F:crabspecgammas}
%\end{center}
\end{figure}

To check the derived efficiency values, an artificial software threshold was applied to
all system data in the following way: 
(a) the 2nd-brightest pixel in each camera must exceed $12\,\mathrm{ph.e.^{*}}$; the
unit $\mathrm{ph.e.^{*}}$ indicates that the pixel amplitudes
were divided by $\kappa'_{\mathrm{tot}}$ beforehand,
according to the image amplitude $amp^{*}_{tel}$;
(b) at least 2 telescopes must surpass this camera threshold. 
By this procedure, the electronic trigger decision of the telescope system was simulated, 
albeit with a sufficiently higher threshold than the hardware value of 6\,ph.e.
This software threshold should provide a homogeneous spectral acceptance for all recorded data.

With the events selected in this manner, three tests were performed:
\\
$\bullet$
The event rate of all system data was redetermined.
The result is shown in Fig.\,\ref{F:ratedevel} upper panel together with the raw event rate
of the system. Within the expected accuracy, the rate is now constant, as expected.
\\
$\bullet$
More important, the $\gamma$-ray acceptance and hence the 
$\gamma$-ray rate for a steady source should also be constant. Figure \ref{F:crabrate}
shows the $\gamma$-ray rate from the Crab nebula and the corresponding background rates for comparison,
obtained from zenith angles below 20\degr,
with and without software threshold.
The Crab nebula is assumed to be a steady TeV emitter on these time scales.
%\cite{} 
Indeed,
the detected $\gamma$-ray rate after software threshold remained within 10\% constant throughout the years.
\\
$\bullet$
Finally, the stability of the spectral acceptance of the detector for $\gamma$-rays was also checked with Crab data.
Figures \ref{F:crabspeccr} and \ref{F:crabspecgammas}
show background and $\gamma$-ray energy spectral ratios from Crab observations.
%%% under low zenith angles.
The spectra were split up into yearly intervals,
and were then divided binwise by the full 1997-2002 spectra (neglecting the correlation).
The histograms were normalized to the respective energy-integrated background event ratios
to correct for the different exposure times.
Data up to 25\degr\ zenith angle were used;
the spectral acceptance for $\gamma$-rays changes quickly within this zenith angle range, but
the compared time intervals have nearly the same zenith angle event distribution
(with the slight exception of period 01-02) which makes the comparison feasible.
A fit $\propto E^{\Delta\alpha}$ (dashed lines) was applied to all ratio histograms to quantify the deviation
from the expected flat ratio (results $\Delta \alpha$ and $\chi^2$ as denoted in the figures),
and also a $\chi^2$ for a constant value ($\chi^2_{\Delta \alpha=0}$) was computed.
Within the given Crab $\gamma$-ray statistics, the spectral $\gamma$-ray ratios are flat,
as expected after the software threshold, and the $\gamma$-ray spectra from different years
are statistically compatible.
For the last period (01-02), a $2\,\sigma$-deviation from the expectation is observed;
the chance probability for the expected flat ratio is 8\%.
%% The reason for 
This deviation is probably caused by the different zenith angle distributions of the compared data,
and is presumably not a hint 
%%% It is therefore not clear whether this is a hint 
towards the start of limitations
% for beginning 
of the scaling procedure for each period. 
Corresponding studies on Crab spectra using the scaled effective areas $A^{*}_{\mathrm{eff}}$
did not show similar deviations from the expectation.

Also the spectral indices for the background spectra agree within $\Delta\alpha = 0.05$;
however, because of the high number of background events, the spectra are not fully compatible,
either reflecting the different zenith angle distributions of the data sets, or
indicating the limitations of the applied rescaling plus software threshold approach.
%The $\chi^{2}$-deviations of the background spectra from
%a flat behaviour are beyond the accuracy which can be obtained here.

Data after application of the discussed software acceptance threshold can be used
to directly compare measured energy spectra of different sources,
obtained at different time periods and with arbitrary further analysis cuts
(e.g.\ \cite{HEGRACasAPaper}). However, in most spectral analyses, the effective areas after
calibration as discussed above were applied to derive energy spectra 
(e.g.\ \cite{Mrk501PaperI,HEGRA1426Letter,KonopelkoCrab}).

\section{Absolute energy threshold}
\label{S:absolute}

To measure $\gamma$-ray spectra and fluxes, one needs to rely on the absolute energy 
calibration of the detector. In effect, besides shower simulations and the atmospheric transmission, 
the total conversion factor between the number of Cherenkov photons and the 
pixel amplitude,
\begin{equation}
  \kappa_{\mathrm{tot}} = \kappa_{\mathrm{el}} \cdot \kappa_{\mathrm{opt}},
\end{equation}
is required for the detector simulation. 

From initial measurements of new detector components, the expected value for the total efficiency 
of a telescope was estimated to be $\kappa_{\mathrm{tot},tel} = 0.12\,\mathrm{FADC}/\mathrm{ph.}$,
where $\kappa_{\mathrm{el},tel}$ was assumed to be 1\,FADC/ph.e.
Besides uncertainties in the derivation of this number, the aging behaviour of the components
is not known {\it a priori}. Hence, an {\it in situ} calibration of the instrument is of course desirable.

The standard method is based on the comparison of the experimentally 
measured trigger rate induced by charged CRs with
the predicted rate derived from 
detailed Monte Carlo simulations \cite{HEGRAPerformance99}. Due to 
uncertainties in the CR flux, spectrum and composition, this
method has an uncertainty of $\approx 15\%$ (or maybe even 22\% \cite{CarstenThesis}) in the energy estimate.

Within the HEGRA experiment,
two alternative methods to directly obtain $\kappa_{\mathrm{tot}}$ were investigated,
a muon ring analysis and an installation of a stabilized laser using a calibrated photodiode as reference.
%The results are discussed in the following two subsections.
The muon ring results are discussed in \S\,\ref{SS:muonruns},
\S\,\ref{SS:conclusionabsolute} summarizes the absolute energy threshold investigations.

\subsection{Muon runs}
\label{SS:muonruns}

\begin{figure}[t]
%\begin{center}
%%% \hc{\includegraphics[width=0.9\columnwidth]{calimages.eps}}
%%% \caption{Left: A typical muon event, as seen by a camera. Dark pixels have an entry above 3 photoelectrons
%%% after pedestal subtraction. Right: Image from a typical calibration light pulse,
%%% using a laser in about 65 meters distance and a calibrated photodiode as reference.
%%% Dark illuminated pixels typically have an amplitude of 100\,ph.e. The outer shape of the image 
%%% traces the somewhat irregular contour of the tesselated mirror, the hole in the center is caused by
%%% the shadow of the camera \cite{AndreaAbsolutKalibration}.
\hc{\includegraphics[width=0.7\columnwidth]{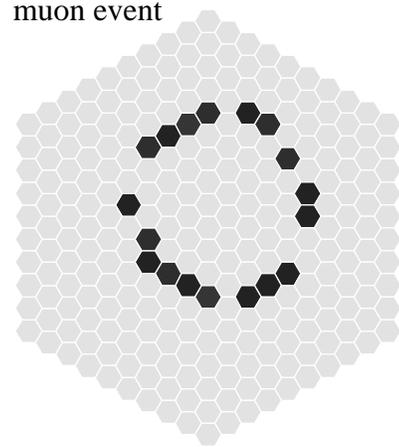}}
\caption{A typical muon event, as seen by a camera. Dark pixels have an entry above 3 photoelectrons
after pedestal subtraction; typically, the pixel amplitudes do not exceed 15 photoelectrons.
%The numbers in the pixels denote the pixel
%amplitudes in photoelectrons, after pedestal subtraction.
}
\label{F:muonring}
%\end{center}
\end{figure}

\begin{figure}[t]
%\begin{center}
\hc{\includegraphics[width=1.0\columnwidth]{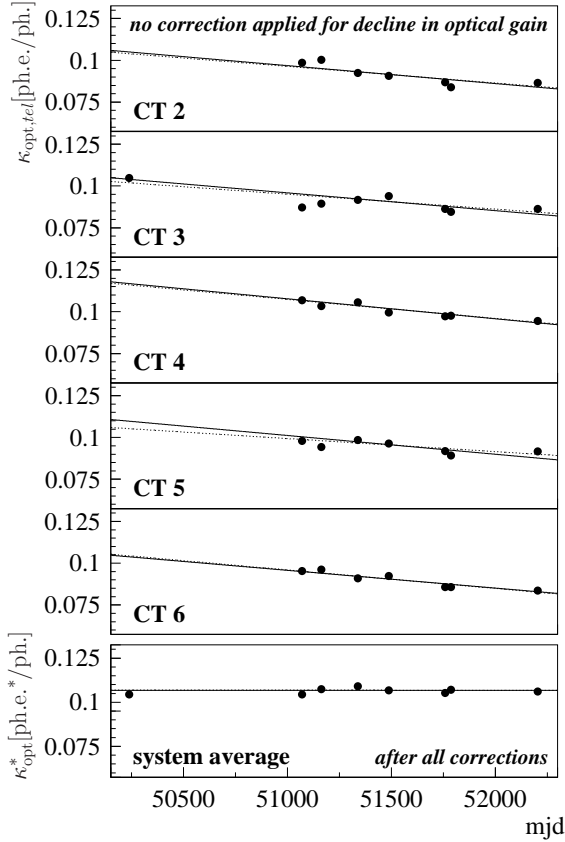}}
\caption{
Measurements of the absolute optical telescope efficiencies.
The data points are results from {\it muon runs}.
%%% Filled circles are results from {\it muon runs}, open circles from the calibrated photodiode measurements.
%The measured values are the absolute efficiencies
%of converting a photon into a signal amplitude, for individual system telescopes. 
In the upper panels, the development of the efficiency vs. time is shown for all telescopes
individually. Here, the signal amplitudes were corrected for the electronic conversion
factor $\kappa_{\mathrm{el},tel}$.
The fits (solid and dotted lines) describe the data very well; details are explained in the text.
%The solid lines are linear fits to the muon results
%with a fixed slope of -3.7\% per year,
%as expected from the rate development. Dashed lines denote fits with fitted slope. Both
%fits are in good agreement, showing that the results of the rate development and of the muon
%ring analysis are well consistent.
In the lower panel, the system average of all telescope
efficiencies is plotted, where the relative gain drop of the optical efficiency 
($\kappa'_{\mathrm{opt}}$)
was also corrected. As expected, this corrected system average is constant in time.
%Results of a muon run analysis, used to determine the absolute optical efficiency. 
%The dates of the first and the last data points correspond 
%to periods 77 and 86 in Fig.\,\ref{F:ratedevel}, respectively.
}
\label{F:muoneff}
%\end{center}
\end{figure}

%When a muon hits the telescope reflector,
%the total number of Cherenkov photons hitting the mirror depends only on event characteristics,
%which can be determined from geometrical properties of the measured muon ring image.
%For such runs, only special trigger conditions (no system trigger, 5 pixels above 6\,mV) are
%required. 

The Cherenkov radiation which is emitted by a muon hitting the telescope reflector or passing nearby
creates a ring-shaped image in the camera (see Fig.\,\ref{F:muonring}) \cite{FleuryMuons,VacantiMuons}.
The total number of photons which have hit the reflector can be reconstructed
from geometrical properties of the ring image alone, i.e.\ from the 
diameter and position of the muon ring and the asymmetry of the ring illumination.
Hence, the measured intensity of the muon ring can be directly used to determine the efficiency
with which the instrument converts photons into pixel amplitudes.

During normal operation of a telescope system like the HEGRA IACT system, 
only very few muons are recorded, since local muons do not release a system trigger.
Hence, special {\it muon runs} are required to effectively trigger on muons. For HEGRA,
a trigger condition of at least 5 pixels, of which at least 2 must be adjacent, above 6\,mV
was used for camera trigger, and the system trigger was disabled.
The analysis procedure is described in \cite{OliverDipl}. In general, the agreement
between simulated muon events and data is good. However, for a HEGRA-sized reflector, even the
highest pixel amplitudes in typical muon rings are close to the trigger threshold. Therefore,
the determined efficiencies are slightly biased  and parameter dependent, which can be explained
by the influence of the night sky background and threshold effects. 
The estimated systematic error for the such derived absolute efficiencies $\kappa_{\mathrm{tot},tel}$
is of the order of 10\%.

The comparison between different efficiencies, derived both for different telescopes and
for different times, is however accurate to the few percent level. 
In Fig.\,\ref{F:muoneff} the telescope efficiencies are drawn as a function of time (filled circles).
Each data point was derived from all {\it muon runs} which were taken within one or two nights.
The statistical errors of the efficiencies are below the marker size. 
In the upper part of the figure, the values are shown for all individual telescopes.
%%% As already mentioned in \S\,\ref{S:stability}, 
The pixel amplitudes were in this case corrected
for the influence of $\kappa_{\mathrm{el},tel}$
(and hence expressed in units of ph.e.).
Therefore, the derived efficiencies are a direct measure of the
absolute optical efficiencies $\kappa_{\mathrm{opt},tel}$. 

A common decline of the optical efficiency is observed for
all telescopes. 
%The solid lines are linear fits to the data, with a fixed slope of -3.7\% per year.
The solid lines have a fixed slope of -3.7\% per year, and were only normalized to the muon results.
This value for $\mathrm{d}\kappa_{\mathrm{opt}}/\mathrm{d}t$ was derived
on average for the system from the trigger rate analysis 
(see Fig.\,\ref{F:ratedevel} and Eq.\,\ref{Eq:optdecline}),
%as explained in \S\,\ref{SS:conversion}, 
and describes the results for the individual telescopes very well.
The difference to fits where the slope was allowed to vary freely (dotted lines) is small,
showing that the results of the rate development and of the muon
ring analysis are well consistent.

Deviations of a few percent of individual data points from the average temporal decline are expected,
as drifts of the electronic conversion factors $\kappa_{\mathrm{el},tel}$ of that order within an
observing period are possible. These drifts are not corrected as only average values of 
$\kappa_{\mathrm{el},tel}$ from within a period are applied to the data.

As one can also see from Fig.\,\ref{F:muoneff}, the difference of the optical efficiencies
between telescopes is about 5\%.
These differences have not entered the standard calibration procedure.
Since in most cases at least 4 telescopes were included in the system, the impact on the energy
scale is negligible, and the influence on the energy resolution is presumably only small for standard
energy reconstructions.
%(See also \cite{HofmannIntercalibration} for a different approach to telescope intercalibration.)

In the lower panel of Fig.\,\ref{F:muoneff}, the system average of the optical efficiencies
$\kappa^{*}_{\mathrm{opt}}$ is plotted,
derived after correction of the pixel amplitudes according to
\begin{equation}
  c^{*}_{pix} = \left(\kappa_{\mathrm{el},tel} \cdot \kappa'_{\mathrm{opt}}\right)^{-1} c_{pix}
\end{equation}
(compare to \S\,\ref{SS:relativecheck}).
As expected, the such corrected system efficiency is constant in time. The solid line is a linear fit with zero slope,
which is in good agreement with the data and with the fit where the slope was allowed to vary (dotted line).

The system average value 
\begin{equation}
  \kappa^{*}_{\mathrm{opt}} [\mathrm{ph.e.^{*}/photon}] = 10.7\%
\end{equation}
is by definition equivalent to the total efficiency $\kappa_{\mathrm{tot}}$
of the reference period, where $\kappa'_{\mathrm{tot}}=1\,\mathrm{FADC/ph.e.^{*}}$.
The measured value agrees within 15\% with the expected value of 12\%, which was used to 
perform the detector simulations.

\subsection{Conclusions for the absolute energy calibration}
\label{SS:conclusionabsolute}

Both methods,
the muon ring calibration as well as the laser illumination using a calibrated photodiode as reference,
have the potential to reduce the error on the energy scale down to a few percent.
The laser method was described in detail in \cite{AndreaAbsolutKalibration};
the result obtained for CT\,4 at an early stage of the system agrees within 10\% 
with the {\it muon run} result.

For the HEGRA system however,
% as discussed above,
systematic effects limit the evaluation of the parameters 
in the muon ring analysis as well as in the laser calibration.
From all performed investigations,
we estimate the uncertainty on the energy scale to be of the order of 15\%.

\section{The long-term stability and sensitivity of the system}
\label{S:stability}

\begin{figure}[t]
%\begin{center}
\hc{\includegraphics[width=1.0\columnwidth]{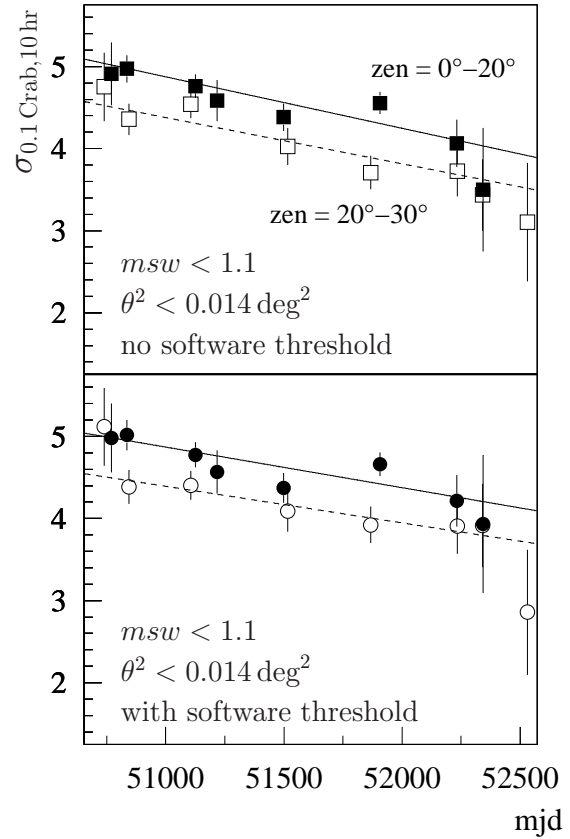}}
\caption{Sensitivity of the HEGRA system for weak sources, derived from Crab data as a function
         of the experiment's lifetime. The values denote the expected significances for a ten hours observation,
	 in units of standard deviations $\sigma$, 
	 for a $\gamma$-ray source which has 10\% of the Crab's source strength.
         Filled symbols are for zenith angles below $20\degr$, open symbols for zenith angles between $20\degr$ and $30\degr$.
         In the upper panel, angular and shape cuts were applied as in a standard analysis approach for weak sources.
	 In the lower panel, an additional software threshold as discussed in the text was applied.
         }
\label{F:crabsigmas}
%\end{center}
\end{figure}

The aging of the HEGRA IACT system components can be estimated
from a linear fit to the time development of the efficiencies 
$\kappa_{\mathrm{el}}$ and $\kappa'_{\mathrm{opt}}$ (see \S\,\ref{S:relative}, Fig.\,\ref{F:ratedevel}).
The main contribution to the aging came from
%was mainly caused by
a decrease of the PM gain in the camera amplification channel of
\begin{equation}
  \mathrm{d}\kappa_{\mathrm{el}}/\mathrm{d}t = (-8.0 \pm 0.3)\% \mathrm{~per~year};
\end{equation}
this is most probably an aging effect of the
PM's last dynode.
The continuous decrease of $\kappa_{\mathrm{opt}}$ 
\begin{equation}
  \mathrm{d}\kappa_{\mathrm{opt}}/\mathrm{d}t = (-3.7 \pm 0.3)\% \mathrm{~per~year}
  \label{Eq:optdecline}
\end{equation}
is presumably caused by a deterioration of the
reflecting layer of the glass mirrors. This was also supported by direct measurements of the
reflectivity of several mirrors, which were performed in August 2000 
(Fig.\,\ref{F:ratedevel} lower panel, star marker at HEGRA observation period 100);
these measurements were performed relative to new mirrors, the result is hence
expressed in the same units as $\kappa'_{\mathrm{opt}}$.

The gradient of the time evolution of $\kappa_{\mathrm{opt}}$ was also 
confirmed by the {\it muon run}
analysis, as discussed in \S\,\ref{S:absolute}. 
%%% For this purpose, muon rings are evaluated using pixel
%%% amplitudes which are corrected for $\kappa_{\mathrm{el}}$.
The results (filled circles in Fig.\,\ref{F:muoneff}, upper panels) are 
%%% a measure of the absolute value of $\kappa_{\mathrm{opt}}$. 
very well described by
the solid lines 
%are fits to the data with
which have a fixed gradient of 
$-3.7\%$ per year.
%%%  and describe the data very well. 
%Details are discussed in the following section.

Finally, in Fig.\,\ref{F:crabsigmas} we show the sensitivity of the HEGRA telescope system (in a 4-telescope-setup),
derived from Crab data, as a function of the detector lifetime. The sensitivity is expressed in units of the expected
standard deviation, when observing a source with 10\% of the Crab's source strength for 10 hours.
The values were calculated with the measured $\gamma$-ray and background rates from Crab, as shown in 
Fig.\,\ref{F:crabrate}.
In the upper panel, standard analysis cuts for the search for weak sources were applied.
To first order, the sensitivity has remained constant. 
However, as can be seen in Fig.\,\ref{F:crabsigmas}, a slight
decrease of the sensitivity with time is observed;
the fits yield 
\begin{equation}
  \mathrm{d}\sigma/\mathrm{d}t = (-4.1 \pm 0.6)\% \mathrm{~per~year}.
\end{equation}
The same trend is visible after application of the software
threshold as discussed in \S\,\ref{SS:relativecheck}, which provides a nearly constant $\gamma$-ray acceptance (Fig.\,\ref{F:crabsigmas}, lower panel, 
compare to Fig.\,\ref{F:crabrate}).
The reason for this behaviour 
is presumably the following: In order to obtain a constant energy threshold
(within a range of 100\,GeV), the electronic gain of the detector was increased in the
long term, to compensate for the losses in mirror reflectivity (cf.\ Fig.\,\ref{F:ratedevel}). 
This however led to a slight deterioration of the background rejection capability
and thus to an increasing background by charged CR events.

\section{Conclusion}

Throughout the years, the HEGRA IACT system performance was well
monitored and found to be stable. In the following, we summarize the
most important technical data of the experiment which were found or confirmed by these
investigations:

\begin{itemize}
  \item nominal energy threshold: \\
        $E_{\mathrm{thr,0}} = 500\,\mathrm{GeV} \pm 15\%(\mathrm{sys})$
  \item actual energy threshold: \\
        $E_{\mathrm{thr}}   = E_{\mathrm{thr,0}} + \Delta E$, $\Delta E \le 100\,\mathrm{GeV}$
  \item aging of the electronic amplification chain (presumably aging of the PM's last dynode):\\
        $\mathrm{d}\kappa_{\mathrm{el}}/\mathrm{d}t = -(8.0 \pm 0.3)\% \mathrm{~per~year}$
  \item aging of the optical components (presumably mirror aging):\\
        $\mathrm{d}\kappa_{\mathrm{opt}}/\mathrm{d}t = -(3.7 \pm 0.3)\% \mathrm{~per~year}$
  \item absolute pointing accuracy:\\
%        $\Delta \vec{x} \le 25\,\mathrm{arcsec}(\mathrm{sys})$
        $\Delta Ra \le 25\,\mathrm{arcsec}(\mathrm{sys})$ \\
        $\Delta Dec \le 25\,\mathrm{arcsec}(\mathrm{sys})$
  \item alignment of the $\gamma$-ray expectation value for the 
        shape parameter mean scaled width with the nominal value of 1:\\
        $\Delta msw_0 \le 2\%(\mathrm{sys})$ \\
        which results in an uncertainty of the tight shape cut efficiency:\\ 
        $\Delta \kappa_{\gamma, msw<1.1} < 1\%(\mathrm{sys}) $
  \item flux sensitivity ($5\,\sigma$, observations at zenith):\\
        $\phi_{\mathrm{min}} = 0.1\,\mathrm{Crab} \times (t/10\,\mathrm{hrs})^{-1/2} $ \\
        with a time gradient of about: \\
        $\mathrm{d}\sigma/\mathrm{d}t = (-4.1 \pm 0.6)\% \mathrm{~per~year}$.
\end{itemize}

We believe that with the HEGRA system, an important contribution has been made to the
effort of establishing ground based imaging Cherenkov telescopes as precision
detectors, well suited within the broad range of astronomical instruments.
HEGRA has pioneered the stereoscopic aspect of the ground based air Cherenkov imaging technique,
a basic feature which is now being realized in most of the next generation instruments.
The experience gained with HEGRA will be 
profitable for the future projects,
such as the H.E.S.S. telescope system which is currently under construction in Namibia.

\section{Acknowledgements}
The support of the German ministry for Research and
technology BMBF and of the Spanish Research Council CICYT is gratefully
acknowledged.
We thank the Instituto de Astrof\'{\i}sica de Canarias
for the use of the site and for supplying excellent working conditions at
La Palma. We gratefully acknowledge the technical support staff of the
Heidelberg, Kiel, Munich, and Yerevan Institutes.

%\begin{thebibliography}{}

\bibliographystyle{elsart-num}
\bibliography{paper}

%\end{thebibliography}

\end{document}